\documentclass[aps,prd,a4paper,12pt,nofootinbib,notitlepage,preprintnumbers]{revtex4-1}

\usepackage{amsmath}   
\usepackage{braket}    
\usepackage{xspace}    
\usepackage{graphicx}  
\usepackage{xcolor}    
\usepackage{hyperref}  
\usepackage{orcidlink} 

\renewcommand{\a}{\mathbf{a}}
\renewcommand{\k}{\mathbf{k}}
\newcommand{\p}{\mathbf{p}}
\newcommand{\q}{\mathbf{q}}
\newcommand{\x}{\mathbf{x}}
\newcommand{\y}{\mathbf{y}}
\newcommand{\n}{\mathbf{n}}
\newcommand{\0}{\mathbf{0}}
\newcommand{\2}{\mathbf{2}}
\newcommand{\3}{\mathbf{3}}

\newcommand{\Ac}{\mathcal{A}}
\newcommand{\Bc}{\mathcal{B}}
\newcommand{\Cc}{\mathcal{C}}
\newcommand{\Gc}{\mathcal{G}}
\newcommand{\Hc}{\mathcal{H}}
\newcommand{\Ic}{\mathcal{I}}
\newcommand{\Kc}{\mathcal{K}}
\newcommand{\Mc}{\mathcal{M}}
\newcommand{\Oc}{\mathcal{O}}
\newcommand{\Pc}{\mathcal{P}}
\newcommand{\Tc}{\mathcal{T}}
\newcommand{\Vc}{\mathcal{V}}
\newcommand{\Yc}{\mathcal{Y}}

\newcommand{\diff}{\mathrm{d}}
\DeclareMathOperator{\im}{Im}
\newcommand{\bs}[1]{\boldsymbol{#1}}
\newcommand{\bh}[1]{\mathbf{\hat{#1}}}
\newcommand{\nn}{\nonumber}

\newcommand{\cf}{cf.\xspace}
\newcommand{\ie}{i.e.\xspace}

\newcommand{\kstarp}{k_p^\star}
\newcommand{\pstark}{p_k^\star}
\newcommand{\hatkstarp}{\bh{\k}_{p}^\star}
\newcommand{\hatpstark}{\bh{\p}_{k}^\star}
\newcommand{\qstark}{q_k^\star}
\newcommand{\qstarp}{q_p^\star}
\newcommand{\jmaxcg}{j_{\textrm{max}}}

\definecolor{wm_green}{HTML}{115740}

\hypersetup{
pdftitle     = {An improved partial-wave projection of the one-particle exchange in relativistic three-body scattering},
pdfsubject   = {Scattering Theory},
pdfkeywords  = {Scattering, Partial Waves, QCD, Hadron, Physics, Lattice},
pdfauthor    = {Nicholas C.~Chambers, Andrew W.~Jackura},
pdfnewwindow = {true},
colorlinks   = {true},
linkcolor    = {wm_green},
citecolor    = {wm_green},
filecolor    = {wm_green},
urlcolor     = {wm_green}
} 

\newcommand{\wm}{Department of Physics, 
William \& Mary, 
Williamsburg, VA 23187, USA}
\newcommand{\jlab}{Theory Center, Thomas  Jefferson  National  Accelerator  Facility, Newport  News,  Virginia  23606,  USA}

\begin{document}

\title{An improved partial-wave projection of the one-particle exchange in relativistic three-body scattering}

\preprint{JLAB-THY-26-4943}


\author{Nicholas~C.~Chambers\orcidlink{0009-0007-4891-9963}}
\email[e-mail:~]{ncchambers@wm.edu}
\affiliation{\wm}
\affiliation{\jlab}

\author{Andrew~W.~Jackura\orcidlink{0000-0002-3249-5410}}
\email[e-mail:~]{awjackura@wm.edu}
\affiliation{\wm}
\affiliation{\jlab}

\begin{abstract}

A critical component of relativistic three-body scattering amplitudes is the one-particle exchange (OPE) process, wherein a single particle is exchanged between incoming and outgoing two-body subsystems. 
We present an improved, finite-sum expression for the partial-wave OPE that is valid for three massive, spinless particles in arbitrary angular momentum configurations. 
A collection of analytic and numerical checks demonstrate that our finite-sum expression reproduces the known properties of the partial-wave OPE, including its threshold scaling behavior and singularity structure. 
Compared with previous work, this result more efficiently handles the rapid proliferation of partial waves contributing to any total angular momentum $J$, enabling the construction of robust wavesets for three-body amplitude analyses.

\end{abstract}

\date{\today}
\maketitle

\section{Introduction}
\label{sec:intro}

Three-body scattering amplitudes are essential for interpreting a broad range of nuclear and hadronic phenomena, particularly resonances that decay into three-particle final states. 
Experimentally, they are constrained through multi-hadron production and decay distributions, where rescattering and kinematic singularities can obscure or mimic resonance signals~\cite{COMPASS:2020yhb}. 
Their theoretical description is complicated by multi-dimensional kinematics and the interplay of coupled two-body subchannels, rescattering via particle exchange, and genuine three-body interactions. 
Decades of progress in reaction theory and amplitude analysis have produced representations consistent with unitarity and analyticity~\cite{Fleming:1964zz,Aaron:1968aoz,Mai:2017vot,Jackura:2019bmu,Mikhasenko:2019vhk,Dawid:2020uhn,Dawid:2023jrj,Jackura:2025wbw,Zhang:2026eez}, while finite-volume methods now provide a path to constrain their dynamics from nonperturbative lattice QCD~\cite{Hansen:2014eka,Hansen:2015zga, Hansen:2019nir,Mai:2021lwb,Blanton:2020gha}.

In lattice QCD, scattering amplitudes are obtained indirectly from the discrete finite-volume spectrum~\cite{Briceno:2017max}. 
For two-particle systems, the L\"uscher formalism and its extensions relate this spectrum to a two-body $\mathcal K$ matrix, which is then used to construct the physical amplitude~\cite{Luscher:1985dn,Luscher:1986n2,Luscher:1990ux,Rummukainen:1995vs,Kim:2005gf,He:2005ey,Davoudi:2011md,Hansen:2012tf,Briceno:2012yi,Briceno:2013lba,Briceno:2014oea,Romero-Lopez:2018zyy}. 
Generalizations to three particles constrain two- and three-body $\mathcal K$ matrices through quantization conditions~\cite{Hansen:2014eka,Hansen:2016ync,Polejaeva:2012ut,Briceno:2012rv,Mai:2017bge,Meng:2017jgx,Briceno:2017tce,Hammer:2017uqm,Hammer:2017kms,Guo:2017ism,Guo:2017crd,Doring:2018xxx,Briceno:2018mlh,Briceno:2018aml,Pang:2019dfe,Blanton:2019igq,Romero-Lopez:2019qrt,Blanton:2020jnm,Blanton:2020gha,Hansen:2020zhy,Blanton:2020gmf,Romero-Lopez:2020rdq,Muller:2020vtt,Muller:2020wjo,Blanton:2021mih,Muller:2022oyw,Draper:2023xvu,Hansen:2025oag,Alotaibi:2025pxz}, after which the physical $\3\to\3$ amplitude is reconstructed by solving integral equations~\cite{Hansen:2015zga,Mai:2017vot,Jackura:2020bsk,Jackura:2022gib}. 
Parallel advances in the finite-volume formalism and $S$-matrix theory~\cite{Guo:2015zqa,Mai:2017vot,Jackura:2018xnx,Mikhasenko:2018bzm,Mikhasenko:2019vhk, Briceno:2019muc, Dawid:2020uhn, Jackura:2019bmu, Jackura:2020bsk, Dawid:2023jrj, Dawid:2023kxu, Jackura:2022gib, Jackura:2023qtp, Feng:2024wyg,Briceno:2024ehy, Jackura:2025wbw} have established this workflow for many three-particle systems,~\footnote{Examples of such systems include three pions~\cite{Hansen:2020zhy}, $K\bar{K}\pi$~\cite{Draper:2024qeh}, and three neutrons~\cite{Draper:2023xvu}.} and lattice determinations of three-hadron amplitudes are now under way~\cite{Blanton:2019vdk,Hansen:2020otl,Blanton:2021llb,Dawid:2024dgy,Yan:2024gwp,Dawid:2025doq,Yan:2025mdm,Briceno:2025yuq,Feng:2026ixm}. 
To date, applications of this framework have focused exclusively on excited hadron spectroscopy.

Spectroscopic studies typically organize amplitudes into sectors of definite spin-parity $J^P$ to isolate resonances with the corresponding quantum numbers. 
The partial-wave projection of a $\3\to\3$ amplitude is more involved than that of its $\2\to\2$ counterpart because of the larger number of kinematic degrees of freedom~\cite{Briceno:2024ehy}. 
Even at fixed $J^P$, the three-body amplitude $\Mc_3^{J^P}$ receives contributions from infinitely many internal angular momentum configurations, so its partial-wave expansion must be truncated in practice. 
Such truncations and their associated systematics are familiar from experimental partial-wave analyses of three-hadron production based on isobar models~\cite{COMPASS:2014vkj,COMPASS:2015gxz,LHCb:2019xmb,LHCb:2019kea,GlueX:2024erj,Kopf:2020yoa}. 
Because of the challenges of the amplitude formalism, early exploratory three-body lattice QCD studies extracted only $\Kc$ matrices or equivalent short-distance quantities~\cite{Beane:2007es,Detmold:2008fn,Blanton:2019vdk,Blanton:2021llb}, while the few subsequent amplitude calculations have been restricted to $J=0$ or $1$ with small wavesets~\cite{Hansen:2015zga,Hansen:2020otl,Jackura:2020bsk,Mai:2021nul,Yan:2024gwp,Yan:2025mdm,Feng:2026ixm}. 
As higher energies are explored, the waveset must be enlarged and the truncation must be varied to quantify the associated systematic uncertainty, as in previous two-body spectroscopy studies~\cite{Dudek:2010ew,Dudek:2012xn,Wilson:2015dqa}. 
Furthermore, studying tensor and higher-spin resonances with three-body decay modes, such as the $a_2(1320)$, requires extending these analyses for higher $J$~\cite{ParticleDataGroup:2026mpi}.

A key feature, and source of complexity, in three-particle reactions is the one-particle-exchange (OPE) mechanism, in which a particle is exchanged between different initial- and final-state two-body subsystems~\cite{Fleming:1964zz,Holman:1965mxx, Kamal:1965vto}. 
The OPE is represented by a purely kinematic function, denoted $\Gc$, whose singular structure arises when the exchanged particle goes on shell. 
It is present even in the minimal construction of $\Mc_3$ consistent with $S$-matrix unitarity and forms a central component of integral equation representations of the $\3\to\3$ amplitude~\cite{Aaron:1968aoz,Mai:2017vot,Jackura:2019bmu,Mikhasenko:2019vhk}. 
For amplitudes of definite $J^P$, this construction requires the corresponding partial-wave projection $\Gc^{J^P}$~\cite{Briceno:2024ehy}.

Reference~\cite{Jackura:2023qtp} presented a general procedure for computing $\Gc^{J^P}$. 
This result has since been used in phenomenological amplitude analyses data~\cite{Winney:2026jja} and in lattice QCD studies~\cite{Dawid:2024dgy,Dawid:2025doq,Briceno:2025yuq,PitangaLachini:2026lyd}. 
In that construction, however, the projected OPE is expressed in terms of coefficients that must be determined separately for each target $J^P$ and each combination of internal angular momenta.~\footnote{Reference~\cite{Jackura:2023qtp} provided a \texttt{Mathematica} notebook to automate the calculation of these coefficients.} 
The procedure must therefore be repeated for every matrix element in the channel of interest, as seen in recent three-body applications that include the OPE across a large waveset~\cite{Dawid:2024dgy,Dawid:2025doq}. 
In Ref.~\cite{Dawid:2025doq}, for example, the calculation involved ten independent spin-orbit subchannels, some requiring as many as six coefficients in a Legendre expansion. 
Consequently, the case-by-case approach becomes increasingly slow and susceptible to transcription errors as the partial-wave truncation is relaxed. 

Here we present an improved formulation of the partial-wave OPE result of Ref.~\cite{Jackura:2023qtp}, obtaining a compact, universal finite-sum representation that is valid for arbitrary internal angular momentum configurations and arbitrary $J^P$. 
The central challenge lies in relating angular functions defined in different rest frames. 
We show how the Lorentz-boosted angular dependence of the two-body subsystems can be explicitly expanded in terms of orthogonal functions defined in the three-body rest frame. 
Standard properties of these functions and $\mathrm{SU}(2)$ couplings then reduce the partial-wave OPE function to a finite-sum representation that is straightforward to implement numerically for $\3\to\3$ scattering of arbitrary spinless particles. 
Although the present result is derived for spinless particles, the underlying projection strategy may provide a useful starting point for generalizations to particles with nonzero intrinsic spin.

The remainder of this article is organized as follows. 
Section~\ref{sec:summary_results} reviews the OPE function and its kinematic inputs, states our main result, and compares it with previous applications. 
This section provides a self-contained guide to applying the result without consulting the remainder of the article. 
Section~\ref{sec:improved_pwp} derives the finite-sum formula in terms of Legendre functions of the second kind, and we obtain the \emph{boost coefficients} that relate the angular functions between frames. 
We also analytically verify the expected momentum scaling in several threshold limits. 
Section~\ref{sec:numerical} presents numerical results for the threshold behavior and energy dependence of the partial-wave OPE, together with the relative magnitudes of the spin-orbit subchannel contributions to selected $J^P$ wavesets. 
We conclude in Sec.~\ref{sec:summary}, and present various supporting technical details in App.~\ref{app:sec:ang_mom}--\ref{app:sec:parity_suppression}.

\section{Summary of Results}
\label{sec:summary_results}

Here, we briefly recapitulate the OPE function and the definitions needed for the improved projection strategy. 
Additionally, we summarize our main results for its partial-wave matrix elements for the interested practitioner. 
We follow the notation, conventions, and definitions of Ref.~\cite{Jackura:2023qtp}, which provides additional background.

\subsection{Recapitulation of the OPE}
\label{sec:recap_ope}

\begin{figure}
    \centering
    \includegraphics[width=0.55\linewidth]{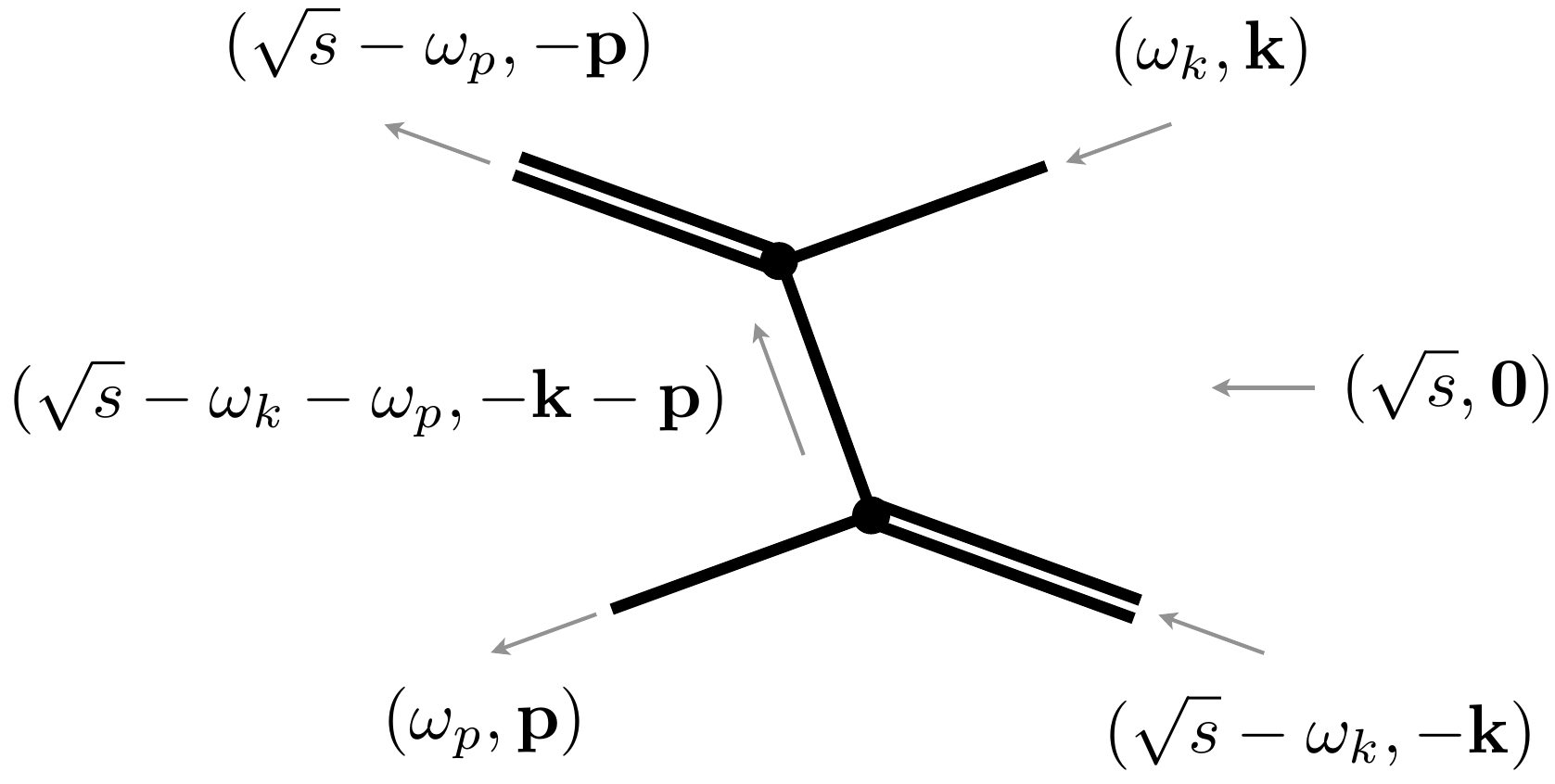}
    \caption{Diagrammatic depiction of the OPE with kinematic assignments used in this work. Single particles are denoted by single lines, while the pair is denoted by the double line.}
    \label{fig:ope_diagram}
\end{figure}

As in Ref.~\cite{Jackura:2023qtp}, we consider a system of three massive, spinless particles.~\footnote{We work in Minkowski spacetime with metric signature $(+---)$ and natural units, $\hbar=c=1$.}
For the OPE process illustrated in Fig.~\ref{fig:ope_diagram}, two incoming particles form a \emph{pair} that separates into the exchanged particle and the final-state \emph{spectator}. 
The exchanged particle then couples to the initial-state spectator to form the final-state pair. 
We denote the initial- and final-state spectator masses by $m_k$ and $m_p$, respectively, and the exchanged-particle mass by $m_e$. 
The product of the intrinsic parities of the three particles is $\eta$. 
We evaluate the system in the total three-body center-of-momentum (CM) frame, where the total four-momentum is $(\sqrt{s},\0)$ and $s$ is the three-body invariant mass squared.
For fixed $s$, the initial-state spectator has four-momentum $(\omega_k,\k)$, while the pair has four-momentum $(\sqrt{s}-\omega_k,-\k)$.~\footnote{Single particle states have the usual relativistic normalization $\braket{\p|\k} = (2\pi)^3\,2\omega_k\,\delta^{(3)}(\p-\k)$.}
Here $\omega_k = \sqrt{m_k^2 + k^2}$ is the on-shell spectator energy, with $k \equiv \sqrt{\k\cdot\k} \ge 0$. 
For physical kinematics, the pair has invariant mass $\sqrt{\sigma_k}$ bounded by $m_p + m_e \le \sqrt{\sigma_k} < \sqrt{s} - m_k$. 
One may use either $k$ or $\sigma_k$, related by $\sigma_k = (\sqrt{s} - \omega_k)^2 - k^2$, or equivalently
\begin{align}
    k 
    = 
    \frac{1}{2\sqrt{s}}\,
    \lambda^{1/2}(s,m_k^2,\sigma_k) \, ,
    \label{eq:cm_momentum}
\end{align}
where $\lambda(x,y,z) = x^2 + y^2 + z^2 - 2(xy + yz + zx)$ is the K\"all\'en triangle function~\cite{byckling1973particle}. 
The final-state spectator and pair kinematics are defined analogously, with $k\leftrightarrow p$. 
We note that the directions of $\p$ and $\k$ are defined with respect to some arbitrary space-fixed coordinate system in the total CM frame.

The exchanged particle is generally off shell, with four-momentum $(\sqrt{s} - \omega_k - \omega_p,-\k-\p)$. 
We define $u_{pk}$ as the corresponding squared momentum transfer, with
\begin{align}
	u_{pk} & \equiv (\sqrt{s} - \omega_k - \omega_p)^2 - (\p+\k)^2 \, , \nn \\[5pt]
	& = u_{pk}^{(0)} - 2pk(1 + \bh{\p}\cdot\bh{\k}) \, ,
	\label{eq:upk_def}
\end{align}
where $u_{pk}^{(0)} \equiv \sigma_k + m_p^2 - 2(\sqrt{s} - \omega_k)\,\omega_p + 2pk$, while $\bh{\k}\equiv \k/k$ and $\bh{\p}\equiv \p/p$ denote the spectator directions~\cite{Jackura:2023qtp}. 
Thus $\bh{\p}\cdot\bh{\k}$ is the cosine of the CM frame scattering angle between the initial- and final-state spectators, which is bounded by $\lvert \bh{\p}\cdot\bh{\k} \rvert \le 1$. 
When the exchanged particle goes on shell, $u_{pk}=m_e^2$, which fixes the cosine of the scattering angle to $\zeta_{pk}$, \ie,
\begin{align}
	\zeta_{pk} 
	& \equiv 
	(\bh{\p}\cdot\bh{\k})\Big\rvert_{u_{pk} = m_e^2} 
	= -1 + \frac{u_{pk}^{(0)} - m_e^2}{2pk} \, .
	\label{eq:recap_zeta}
\end{align}

The OPE function captures the kinematic singularity associated with one particle going on shell between the incoming and outgoing pairs. 
In the pair-spectator basis this appears as a $u$-channel pole in the quasi-two-body scattering amplitude. 
We use the relativistic covariant form studied in Ref.~\cite{Jackura:2023qtp},
\begin{align}
	\Gc_{S'\lambda',S\lambda}(\p,\k) 
	& = 
	\left(\frac{\kstarp}{\qstarp}\right)^{S'} 
	\frac{4\pi\, 
	Y^*_{S'\lambda'}(\hatkstarp) \, 
	Y_{S\lambda}(\hatpstark)
	}{u_{pk} - m_e^2 + i\epsilon} \, 
	\left(\frac{\pstark}{\qstark}\right)^S \,,
	\label{eq:recap_ope_def}
\end{align}
where the $s$ dependence is implicit and the pole prescription is $\epsilon\to0^+$.~\footnote{The spherical harmonics are defined such that $Y_{\ell m}(\bh{\n}) = Y_{\ell m}(\theta_n,\varphi_n)$ where $\theta_n$ and $\varphi_n$ are the polar and azimuthal angles of the direction vector $\bh{\n}$ resolved with respect to the quantization axis.}
The OPE is a matrix in the initial- and final-pair angular momenta. 
Since the particles are spinless, $S$ and $S'$ are pair orbital angular momenta for the initial- and final-state, respectively, with helicity labels $\lambda$ and $\lambda'$ satisfying $-S\leq\lambda\leq S$ and $-S'\leq\lambda'\leq S'$. 
The quantization axes are chosen along the pair momenta, equivalently opposite to the corresponding spectator directions in the three-body CM frame. 
Thus the angular argument of $Y_{S\lambda}(\hatpstark)$ is resolved with respect to the $-\bh{\k}$ axis, while that of $Y_{S'\lambda'}^*(\hatkstarp)$ is with respect to $-\bh{\p}$.~\footnote{The $^*$ on $Y_{S'\lambda'}^*(\hatkstarp)$ denotes complex conjugation.} 
The spherical harmonics therefore describe the pair formation and decay vertices in their respective pair rest frames, while the barrier factors remove kinematic threshold singularities.

Consider first $\p_k^\star$, the final spectator momentum evaluated in the initial-pair rest frame. 
Its magnitude is
\begin{align}
	\pstark &= \frac{1}{2\sqrt{\sigma_k}}\lambda^{1/2}(\sigma_k,u_{pk},m_p^2) \, .
\end{align}
Away from the exchange pole, this magnitude inherits angular dependence through $u_{pk}=u_{pk}(\bh{\p}\cdot\bh{\k})$. 
At the pole, $u_{pk}=m_e^2$, it reduces to the on-shell breakup momentum 
\begin{align}
	\qstark &= \frac{1}{2\sqrt{\sigma_k}}\lambda^{1/2}(\sigma_k,m_e^2,m_p^2) \, .
\end{align}
The relation between the CM frame momentum and the initial-pair rest-frame momentum is obtained by boosting with velocity $\bs{\beta}_k=-\k/(\sqrt{s}-\omega_k)$,
\begin{subequations}
\begin{align}
	\label{eq:recap_boost_par}
	(\p_{k}^{\star})_{\parallel} & = \gamma_k(\p_{\parallel} - \bs{\beta}_k \,\omega_p) \, , \\[5pt]
	\label{eq:recap_boost_perp}
	(\p_{k}^{\star})_{\perp} & = \p_{\perp} \, ,
\end{align}
\end{subequations}
where $\gamma_k = 1/\sqrt{1-\beta_k^2} = (\sqrt{s} - \omega_k) / \sqrt{\sigma_k}$ and the parallel and perpendicular components are defined with respect to the boost velocity, that is $\p_{\parallel} \equiv (\p\cdot\bh{\bs{\beta}}_k)\,\bh{\bs{\beta}}_k$ and $\p_\perp \equiv\p-\p_\parallel$, and similarly for the components of $\p_k^{\star}$. 
The final-pair rest-frame quantities $\k_p^\star$, $\hatkstarp$, and $\qstarp$ are obtained by the interchange $\p\leftrightarrow\k$.

\subsection{Summary of partial-wave projection result}
\label{sec:summary_ope}

The main result of this work is a compact, finite-sum representation for the partial-wave projection of the OPE defined in Eq.~\eqref{eq:recap_ope_def}. 
Here, we collect the formulae needed to apply this result, leaving the derivation and additional properties to Sec.~\ref{sec:improved_pwp}. 
Since our applications of interests involve systems with definite total angular momentum and parity, $J^P$, we express the result in the spin-orbit basis. 
For a fixed $J^P$ sector, this requires all allowed transitions
${}^{2S+1}L_J \to {}^{2S'+1}L'_J$, where $L$ is the orbital angular momentum between the initial-state pair and its spectator, and $L'$ is defined analogously for the final state. 
For given $S$, $S'$, and $J$, angular-momentum coupling requires $\lvert J-S\rvert \le L \le J+S$ and $\lvert J-S'\rvert \le L' \le J+S'$. 
The target parity further restricts the allowed waves through
$P = \eta(-1)^{L+S} = \eta(-1)^{L'+S'}$.

With the partial-wave expansion normalized as in Ref.~\cite{Jackura:2023qtp}, we show in Sec.~\ref{sec:derivation} that the spin-orbit matrix elements of the OPE take the finite-sum form
\begin{align}
	\Gc^{J^P}_{L'S',\, LS}(p,k)
    &=
    \frac{1}{2pk}
	\sum_{j = 0}^{\jmaxcg}(2j+1) \,
	Q_j(\zeta_{pk} + i\epsilon) \,
	\Hc_{L'S',\, LS}^{j,J^P}(p,k)
	\label{eq:ope_result}
\end{align}
where $Q_j$ is the Legendre function of the second kind of degree $j$, evaluated at the exchange singularity variable $\zeta_{pk}$ defined in Eq.~\eqref{eq:recap_zeta}. 
The logarithmic singularities generated by the angular integration over the exchange pole are captured exclusively within the $Q_j$ functions~\cite{Abramowitz_Stegun}. 
We take the branch cut of $Q_j$ to lie along $\zeta_{pk}\in[-1,1]$, which corresponds to the physical OPE region. 
The $+i\epsilon$ inherited from the covariant propagator specifies the boundary value $Q_j(\zeta_{pk}+i\epsilon)$ as $\epsilon\to0^+$ on this cut. 
The sum over $j$ terminates at the maximum upper-bound $\jmaxcg\equiv J+S'+S$, although individual terms may vanish by angular-momentum selection rules. 
If desired, Eq.~\eqref{eq:ope_result} can be further reduced to the form $\Gc^{J^P} = \Kc_{\Gc}^{J^P} + \Tc^{J^P}Q_0(\zeta_{pk})$ presented in Ref.~\cite{Jackura:2023qtp}, as shown in Sec.~\ref{sec:derivation}.

The function $\Hc^{j,J^P}$ is a matrix in spin-orbit channel space and contains the remaining angular-momentum dependence. 
Its matrix elements are
\begin{align}
	\Hc_{L'S',\, LS}^{j,J^P}(p,k)
	=
	\sum_{\ell' = 0}^{S'} \,
	\sum_{\ell  = 0}^{S } \,
	\sum_{n=\lvert\ell'-\ell \rvert}^{\ell'+\ell}
	(-1)^{n} \,
    \Vc^{n,j,J}_{L'S'}(k,\ell';p,\ell) \,
	\Vc^{n,j,J}_{LS}(p,\ell;k,\ell') \, ,
	\label{eq:ope_HLS}
\end{align}
where the $\Vc$ factors describe the angular and kinematic recoupling at each vertex,
\begin{align}
    \Vc^{n,j,J}_{LS}(p,\ell;k,\ell')
    &\equiv
    \sqrt{\frac{(2L+1)(2\ell+1)}{(2J+1)(2n+1)}}\,
    \left(\frac{p}{q_k^\star}\right)^{S} 
    \sum_{\lambda = -\lambda_{\max}}^{\lambda_{\max}} 
    \Bc^{\ell}_{S\lambda}(p;k)\,
    \Cc^{J\lambda}_{L0,\,S\lambda} \,
    \Cc^{n\lambda}_{J\lambda,\,j0} \,
    \Cc^{n\lambda}_{\ell\lambda,\,\ell'0} \, .
    \label{eq:ope_V}
\end{align}
The ordering of the arguments in $\Vc^{n,j,J}_{LS}(p,\ell;k,\ell')$ is an important part of the definition. 
The pair $(p,\ell)$ labels the boosted vertex, with $\ell$ entering the boost coefficient, while $(k,\ell')$ labels the recoupled spectator side. 
The helicity sum is bounded by $\lambda_{\max}=\min(J,S)$, with the analogous final-state factor using $\lambda'_{\max}=\min(J,S')$, while the SU(2) Clebsch-Gordan coefficients,
$\Cc^{jm}_{j_1m_1,\,j_2m_2}\equiv\braket{jm|j_1m_1,\,j_2m_2}$,  impose the usual angular-momentum selection rules, including
$\lvert\ell-\ell'\rvert\leq n\leq \ell+\ell'$ and
$\lvert n-J\rvert\leq j\leq n+J$.

The remaining ingredients are the \emph{boost coefficients} $\Bc^\ell_{S\lambda}$, which relate the angular dependence of the pair formation/decay vertices, naturally defined in pair rest frames, to the total three-body CM frame. 
They are defined as angular overlaps of spherical harmonics evaluated in different reference frames. 
In Sec.~\ref{sec:boost_coefficients}, we give their integral definition and show how to evaluate them analytically for arbitrary quantum numbers. 
The result is most compactly represented by the recurrence relation
\begin{align}
	\Bc^{\ell}_{S\lambda}(p;k) 
	&=  
	\sum_{a=0}^{S-1}\sum_{b=0}^{1} 
    \sqrt{\frac{(2S+1)(2a+1)(2b+1)}{3S(2\ell + 1)}}
    \nn \\[5pt]
    &\qquad \qquad \times 
    \sum_{\mu_a = -a}^{a} \sum_{\mu_b = -b}^{b} 
    \Cc^{S \lambda}_{S-1, \mu_a,\, 1\mu_b} \,
    \Cc^{\ell 0}_{a0, \, b0} \,
    \Cc^{\ell \lambda}_{a\mu_a,\, b\mu_b}  \,
    \Bc^{a}_{S-1, \mu_a}(p; k) \,
    \Bc^{b}_{1\mu_b}(p; k) \, ,
    \label{eq:boost_coeff_recursive}
\end{align}
with initial conditions $\Bc_{1\lambda}^{1}(p;k)=\delta_{\lambda,1}+\delta_{\lambda,-1}+\gamma_k\delta_{\lambda0}$ and $\Bc_{1\lambda}^{0}(p;k)=-\sqrt{3}\,\delta_{\lambda0}\gamma_k\beta_k\omega_p/p$, where $\delta_{nm}$ is the usual Kronecker delta. 
For $S=0$, one trivially has $\Bc_{0\lambda}^{\ell}(p;k)=\delta_{\ell0}\delta_{\lambda0}$. 
Again, the order of the arguments is part of the definition as $p$ labels the momentum being expressed in the rest frame associated with spectator momentum $k$. 
In Sec.~\ref{sec:boost_coefficients}, we also give an explicit $S$-fold product representation in terms of the elementary $\Bc_{1\lambda}^{\ell}$ coefficients.

Together, Eqs.~\eqref{eq:ope_result}--\eqref{eq:boost_coeff_recursive} constitute the main result of this work. 
They reduce the spin-orbit partial-wave OPE to finite angular-momentum sums, with all nontrivial frame dependence contained in recursively generated boost coefficients. 
This gives an implementation-ready expression for arbitrary target quantum numbers and allowed spin-orbit transitions, replacing the case-by-case construction of Ref.~\cite{Jackura:2023qtp}.

\subsection{Comment on practical applications}
\label{sec:comment_applications}

Compared with Ref.~\cite{Jackura:2023qtp}, this representation is substantially easier to deploy in practical three-body calculations, including both lattice QCD amplitude analyses and phenomenological studies of experimental data. 
As in Ref.~\cite{Jackura:2023qtp}, the nonanalytic dependence associated with the exchange pole is isolated explicitly in the Legendre functions $Q_j$, giving direct control over the analytic structure of the OPE. 
We have verified that the finite-sum representation reproduces all formulae reported in Ref.~\cite{Jackura:2023qtp}, while avoiding the need for a separate construction in each spin-orbit channel.

The authors of Ref.~\cite{Dawid:2025doq} have also studied the partial-wave projection of the OPE for amplitude reconstruction in maximal-isospin $3\pi^+$, $3K^+$, $\pi^+\pi^+K^+$, and $K^+K^+\pi^+$ systems with $J^P=0^-$, $1^+$, and $2^-$. 
They independently computed and tabulated the projections for each relevant ${}^{2S+1}L_J \to {}^{2S'+1}L'_J$ channel. 
We have checked that our finite-sum result reproduces each of these projections, again without performing separate angular integrations for every scattering channel.~\footnote{In correspondence with the authors of Ref.~\cite{Dawid:2025doq}, we confirmed that one reported expression, Eq.~(D29), contains a typographical error, although this does not affect their numerical results. 
Such minor discrepancies are not unexpected in case-by-case calculations of this kind, and further motivate a compact formulation that reduces the number of opportunities for such errors to arise.}
Thus, Eqs.~\eqref{eq:ope_result}--\eqref{eq:boost_coeff_recursive} provide a practical route to studying three-body systems with arbitrary $J^P$, which is critical for hadron spectroscopy. 
Additional numerical checks are presented in Sec.~\ref{sec:numerical}.

\section{Improved partial-wave Projection}
\label{sec:improved_pwp}

As emphasized in Ref.~\cite{Jackura:2023qtp}, the partial-wave projection of the OPE is most naturally carried out in two stages. 
We first project onto definite total angular momentum $J$ in the helicity basis, and then perform a unitary transformation to spin-orbit states of definite $J^P$. 
This organization is useful because the pair angular momentum projections are helicity quantum numbers, which are preserved by the helicity-frame boosts relating the total CM frame to the pair CM frames.

Here we summarize some essential formulae from Ref.~\cite{Jackura:2023qtp}. 
The helicity basis partial-wave projection is given by~\cite{Jacob:1959at,Martin_Spearman}
\begin{align}
	\Gc^J_{S'\lambda',\, S\lambda}(p,k)
	=
	\frac{1}{(4\pi)^2}
	\sum_{m_J=-J}^{J}
	\int \!
	\diff\bh{\p}\, 
	\int \!
	\diff\bh{\k}\,
	D^{(J)}_{m_J\lambda'}(-\bh{\p})\,
	\Gc_{S'\lambda',\, S\lambda}(\p,\k)\,
	D^{(J)*}_{m_J\lambda}(-\bh{\k}) \, ,
	\label{eq:pw_hel_proj}
\end{align}
where $\diff\bh{\k}\equiv \diff\varphi_k\,\diff\cos\theta_k$, and both angular integrals run over the full solid angle. 
The arguments of the Wigner $D$ matrices~\footnote{The Wigner $D$ matrix element is defined such that $D_{m'm}^{(j)}(\bh{\n}) \equiv D_{m'm}^{(j)}(\varphi_n,\theta_n,0)$ where $\theta_n$ and $\varphi_n$ are the polar and azimuthal angles of the direction vector $\bh{\n}$. Properties of the Wigner $D$ matrices and Clebsch-Gordan coefficients used in this work can be found in App.~\ref{app:sec:ang_mom}.}
are the pair directions, equivalently the negative spectator directions. 
Thus
$D_{m_J\lambda}^{(J)}(-\bh{\k}) =
D_{m_J\lambda}^{(J)}(\pi+\varphi_k,\pi-\theta_k,0)$,
where $\theta_k$ and $\varphi_k$ are the spectator polar and azimuthal angles in the space-fixed frame. 
For integer $J$, the symmetry property~\cite{VMK}
\begin{align}
	D_{m_J \lambda}^{(J)}(-\bh{\k}) = (-1)^{J}D_{m_J ,{-\lambda}}^{(J)}(\bh{\k}) \, ,
	\label{eq:wigner_d_symmetry}
\end{align}
allows the projection to be written directly in terms of $\bh{\p}$ and $\bh{\k}$.

We use a standard unitary change of basis from helicity to spin-orbit states,
\begin{align}
    \Gc^{J^P}_{L'S',\, LS}(p,k) 
    & = 
    \sum_{\lambda' = -\lambda'_{\max}}^{\lambda'_{\max}} 
    \sum_{\lambda = -\lambda_{\max}}^{\lambda_{\max}} 
    \Pc_{\lambda'}({}^{2S'+1}L'_J) \,
    \Gc^J_{S'\lambda',\, S\lambda}(p,k) \,
    \Pc_\lambda({}^{2S+1}L_J) \, .
    \label{eq:ope_hel_to_pw}
\end{align}
The sums over helicity indices are bounded by $\lambda_{\max} = \min(J, S)$ and $\lambda'_{\max} = \min(J, S')$. 
We have also introduced $\Pc_\lambda({}^{2S+1}L_J)$, which is the helicity-to-spin-orbit coupling coefficient,
\begin{align}
    \Pc_\lambda({}^{2S+1}L_J) 
    = 
    \sqrt{\frac{2L+1}{2J+1}} \, \Cc^{J \lambda}_{L0, \, S\lambda} \, .
    \label{eq:spin_orbit_coupling_def}
\end{align}
For a fixed $J^P$, $S$, and $S'$, angular-momentum coupling requires
$\lvert J-S\rvert\leq L\leq J+S$ and
$\lvert J-S'\rvert\leq L'\leq J+S'$, while parity imposes
$P=\eta(-1)^{L+S}=\eta(-1)^{L'+S'}$. 
The spin-orbit projected OPE must also exhibit the threshold behavior
$\Gc^{J^P}_{L'S',LS}(p,k)=\Oc(p^{L'}k^L)$
as $p,k\to0$ with $\sigma_p$ and $\sigma_k$ held fixed above their respective thresholds~\cite{Jackura:2023qtp,Briceno:2024ehy}. 
This provides an important check on the finite-sum result derived below, whose threshold behavior is analyzed in Sec.~\ref{sec:threshold} and tested numerically in Sec.~\ref{sec:numerical}.

\subsection{Derivation of the OPE partial-wave projection}
\label{sec:derivation}

To evaluate the partial-wave projection, we first separate the angular dependence on $\bh{\p}$ and $\bh{\k}$. 
The pole term of Eq.~\eqref{eq:recap_ope_def} is resolved with the Neumann expansion~\cite{Abramowitz_Stegun},
\begin{align} 
	\label{eq:neumann_mom_exch}
    \frac{1}{u_{pk} - m_e^2} 
    = 
    \frac{1}{2pk(\zeta_{pk} - \bh{\p}\cdot\bh{\k})} 
    = \frac{1}{2pk}
    \sum_{j=0}^\infty 
    (2j+1) \,
    Q_j(\zeta_{pk}) \,
    P_j(\bh{\p}\cdot\bh{\k}) \, ,
\end{align}
where the $+i\epsilon$ shift is understood through $m_e^2\to m_e^2-i\epsilon$, or equivalently through the boundary value $\zeta_{pk}\to\zeta_{pk}+i\epsilon$, with $\zeta_{pk}$ defined in Eq.~\eqref{eq:recap_zeta}. 
We separate the dependence on $\bh{\p}$ and $\bh{\k}$ in the Legendre polynomials $P_j$ by using its addition theorem in terms of the Wigner $D$ matrices~\cite{VMK},
\begin{align}
	P_j(\bh{\p}\cdot\bh{\k}) 
	& = 
	\sum_{m_j = -j}^{j}
	D_{m_j 0}^{(j)\,*}(\bh{\p}) \, 
	D_{m_j 0}^{(j)}(\bh{\k}) \, .
\end{align}

The nontrivial part of the derivation is the numerator of Eq.~\eqref{eq:recap_ope_def}, since the directions $\hatpstark$ and $\hatkstarp$ are defined in pair rest frames rather than in the total CM frame. 
In Ref.~\cite{Jackura:2023qtp}, this dependence was isolated algorithmically through functions of $\bh{\p}\cdot\bh{\k}$. 
Instead, we exploit the fact that the combination
$(p_k^\star)^S \, Y_{S\lambda}(\hatpstark)$ is a regular solid harmonic evaluated at the boosted spectator momentum $\p_k^\star$.~\footnote{A regular solid harmonic $\Yc_{\ell m}(\x)$ is a homogeneous
polynomial of degree $\ell$ in the Cartesian components of $\x$~\cite{STEINBORN19731}.}
For later convenience, we therefore define
\begin{align}
    \Yc_{S\lambda}(\p_k^\star)
    \equiv
    (p_k^\star)^S \, Y_{S\lambda}(\hatpstark) \, .
\end{align}
This representation allows the numerator to be expanded directly in
spherical harmonics of the spectator directions in the CM frame.

Since $\p_k^\star$ is obtained from $\p$ by a boost along the helicity axis $-\bh{\k}$, its azimuthal angle about this axis is unchanged, \cf~Eqs.~\eqref{eq:recap_boost_par} and~\eqref{eq:recap_boost_perp}. The regular solid harmonic therefore admits the finite expansion
\begin{align}
	\Yc_{S\lambda}(\p_k^\star)
	= 
	p^S \sum_{\ell = 0}^{S} 
	\Bc_{S\lambda}^{\ell}(p;k) \, 
	Y_{\ell \lambda}(\bh{\p}') \, ,
	\label{eq:solid_harmonic_expansion}
\end{align}
where the boost coefficients $\Bc^\ell_{S\lambda}(p;k)$ encode the Lorentz transformation from the total CM frame to the rest frame of the pair recoiling against spectator $k$. The ordering of the arguments is important as $\Bc(p;k)$ describes the boost of momentum $p$ to the pair-$k$ rest frame, while $\Bc(k;p)$ denotes the analogous final-state coefficient. The expansion is finite, with support only for $\ell\leq S$. This follows from the polynomial degree $S$ of the regular solid harmonic and from the fact that the boost along the quantization axis is linear in the CM-frame momentum components. The absence of helicity mixing follows from the same geometry. These properties are proven explicitly in App.~\ref{app:sec:boost_coeff}.

\begin{figure}
    \centering
    \includegraphics[width=0.5\linewidth]{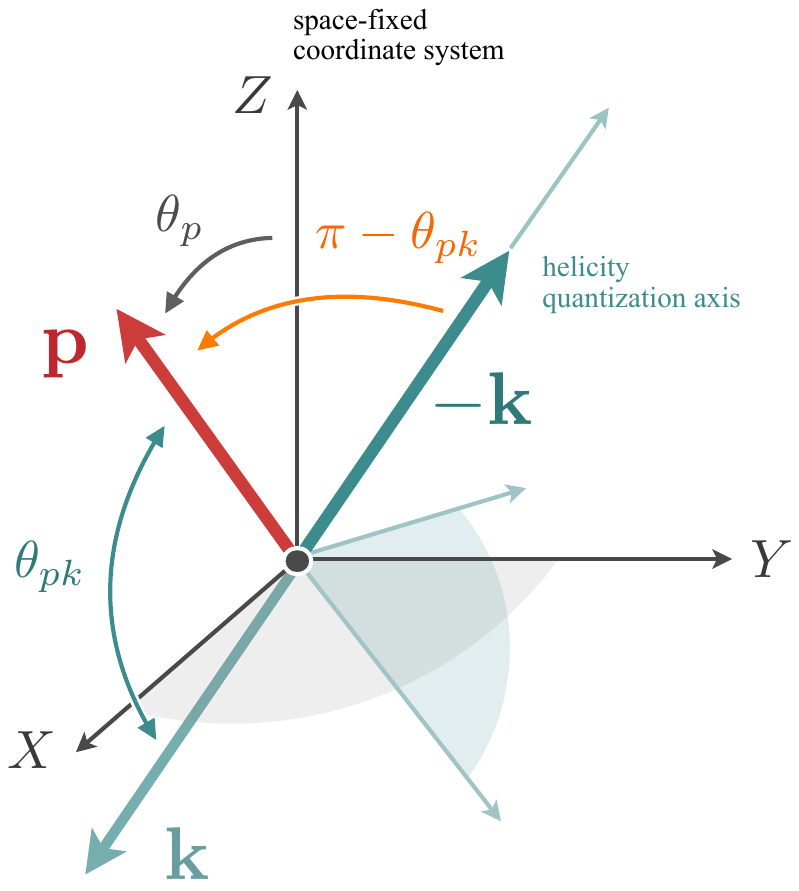}
    \caption{Cartoon illustrating the orientation of $\p$ with respect to the space-fixed coordinate system ($XYZ$) and a coordinate system fixed to the quantization axis, $-\k$. The polar angle of $\p$ is $\theta_p$ with respect to $Z$-axis, while $\pi-\theta_{pk}$ is the polar angle with respect to the helicity quantization axis. The CM frame scattering angle $\theta_{pk}$ is defined through $\cos\theta_{pk} \equiv \bh{\p}\cdot\bh{\k}$.}
    \label{fig:momenta_frames}
\end{figure}
To use Eq.~\eqref{eq:solid_harmonic_expansion} inside the partial-wave projection, we must rotate its spherical harmonics from the helicity frame to the space-fixed frame. In Eq.~\eqref{eq:solid_harmonic_expansion}, the angles of $\p$ are measured relative to the helicity axis $-\bh{\k}$, while the projection in Eq.~\eqref{eq:pw_hel_proj} integrates over $\bh{\p}$ and $\bh{\k}$ in a common space-fixed frame. To distinguish these  two frames, we define $\bh{\p}'$ as the orientation of $\p$ relative to the helicity axis. Figure~\ref{fig:momenta_frames} illustrates this relation. We perform the rotation using
\begin{align}
	Y_{\ell \lambda}(\bh{\p}') 
	& = 
	\sum_{m_{\ell} = -\ell}^{\ell} 
	D_{m_{\ell} \lambda}^{(\ell)}(-\bh{\k}) \, 
	Y_{\ell m_{\ell}}(\bh{\p}) \, , 
	\nn \\[5pt]
	& = (-1)^{\ell}
	\sum_{m_{\ell} = -\ell}^{\ell} 
	D_{m_{\ell}, -\lambda}^{(\ell)}(\bh{\k}) \, 
	Y_{\ell m_{\ell}}(\bh{\p}) \, .
	\label{eq:Y_helicity_to_space_frame}
\end{align}
where in the second line we used the symmetry relation Eq.~\eqref{eq:wigner_d_symmetry}. 

The angles $\bh{\p}'$ are measured with respect to the helicity axis $-\bh{\k}$ on the left-hand side of Eq.~\eqref{eq:Y_helicity_to_space_frame}, while on the right-hand side they are measured in the space-fixed frame. Here $D(-\bh{\k})$ is the unitary transformation from the quantization axis to the final position in the space-fixed frame. 
Combining Eqs.~\eqref{eq:solid_harmonic_expansion} and~\eqref{eq:Y_helicity_to_space_frame} gives the result
\begin{align}
	\sqrt{4\pi} \, \Yc_{S\lambda}(\p_k^\star)
	&  = 
	p^S 
	\sum_{\ell = 0}^{S} 
	(-1)^{\ell}\sqrt{2\ell +1} \,
	\Bc_{S\lambda}^{\ell}(p;k)\, 
	\sum_{m_{\ell} = -\ell}^{\ell} 
	D_{m_{\ell},-\lambda}^{(\ell)}(\bh{\k}) \, 
	D_{m_{\ell} 0}^{(\ell)\,*}(\bh{\p}) \, .
	\label{eq:numerator}
\end{align}
Here, we used the relation Eq.~\eqref{eq:app_ylm_to_D} to convert the spherical harmonics in $\ell$ and $\ell'$ occurring in Eq.~\eqref{eq:Y_helicity_to_space_frame} to the corresponding Wigner $D$ matrices. 
Although the boost coefficients are isolated to their respective pair quantum numbers, the mixing of $\bh{k}$ and $\bh{p}$ in Eq.~\eqref{eq:numerator} prevents the partial-wave OPE matrix elements from cleanly factorizing between initial- and final-state degrees of freedom.
The solid harmonic $\Yc^*_{S'\lambda'}(\k_p^\star)$ has a similar expansion, with $\p\leftrightarrow \k$ in Eq.~\eqref{eq:numerator}.

Inserting Eqs.~\eqref{eq:neumann_mom_exch} and~\eqref{eq:numerator}, together with the analogous final-state expansion, into Eq.~\eqref{eq:recap_ope_def}, and then applying the helicity projection Eq.~\eqref{eq:pw_hel_proj}, gives
\begin{align}
	\Gc_{S'\lambda',\, S\lambda}^{J}(p,k)
	& =
	\frac{1}{2pk} 
	\sum_{j=0}^{\infty} (2j+1) \,
	Q_j(\zeta_{pk}) \, 
	\Hc^{j,J}_{S'\lambda',\, S\lambda}(p,k) \, ,
	\label{eq:ope_pw_helicity}
\end{align}
where we have defined $\Hc^{j,J}_{S'\lambda',S\lambda}$ to capture all the angular momentum dependence,~\footnote{Note that the $\Hc$ function in this work is different from the one introduced in Ref.~\cite{Jackura:2023qtp}.
}
\begin{align}
	\Hc^{j,J}_{S'\lambda',\, S\lambda}(p,k) 
	& = 
	\left(\frac{k}{q_p^\star}\right)^{S'}
	\sum_{\ell'=0}^{S'} 
	(-1)^{\ell'}\sqrt{2\ell'+1} \,
	\Bc_{S'\lambda'}^{\ell'}(k;p) \, 
	\nn \\[5pt]
	& \qquad \times
	\left(\frac{p}{q_k^\star}\right)^{S}
	\sum_{\ell=0}^{S} 
	(-1)^{\ell}\sqrt{2\ell+1} \,
	\Bc_{S\lambda}^{\ell}(p;k) \,
	\nn \\[5pt]
	& \qquad \qquad \times 
	\sum_{m_J=-J}^{J}\sum_{m_j = -j}^{j}
	\sum_{m_{\ell'}= -\ell'}^{\ell'}\sum_{m_\ell = -\ell}^{\ell} 
	\left(\Ic^{J j \ell\ell'}_{m_J m_j m_{\ell} m_{\ell'}}(\lambda')\right)^*
	\Ic^{J j \ell'\ell}_{m_J m_j m_{\ell'} m_{\ell} }(\lambda) \, .
	\label{eq:H_spin_helicity}
\end{align}
Here we have further defined $\Ic$ as the integral over the angular degrees of freedom, 
\begin{align}
	\Ic^{J j \ell'\ell}_{m_J m_j m_{\ell'} m_{\ell}}(\lambda)
	\equiv 
	\frac{1}{4\pi}\,\int \!
	\diff\bh{\k}\,
	D^{(J)\,*}_{m_J,-\lambda}(\bh{\k}) \, 
	D_{m_{\ell}, -\lambda}^{(\ell)}(\bh{\k}) \,
	D_{m_{\ell'} 0}^{(\ell')}(\bh{\k}) \, 
	D_{m_j 0}^{(j)}(\bh{\k}) \, .
	\label{eq:angular_integral}
\end{align}
Here we distinguish $\lambda$ as a quantum number which is not summed over. 
Note the order of indices on the conjugate integral, which is from integrations over $\bh{\p}$, compared with the unconjugated integral, which is due to integrations over $\bh{\k}$.

The angular integral can readily be evaluated using properties summarized in App.~\ref{app:sec:ang_mom}. 
Using the integral over four Wigner $D$ matrices given in Eq.~\eqref{eq:app_cg_four_D_integral},~\footnote{The required second-index condition of Eq.~\eqref{eq:app_cg_four_D_integral} is satisfied in Eq.~\eqref{eq:angular_integral} since $-\lambda=-\lambda+0+0$.} along with the Clebsch-Gordan symmetries Eqs.~\eqref{eq:app_cg_sym_1} and~\eqref{eq:app_cg_sym_2} to convert the helicity labels to the sign convention used below, we find the result
\begin{align}
	\Ic^{J j \ell'\ell}_{m_J m_j m_{\ell'} m_{\ell}}(\lambda) 
	& = 
		\frac{1}{2J+1}
		\sum_{n,m_n}
		(-1)^{\ell'+\ell +j - J} \,
		\Cc^{n m_n}_{\ell m_{\ell},\,\ell'm_{\ell'}}\,
		\Cc^{n\lambda}_{\ell\lambda,\,\ell'0}\,
		\Cc^{Jm_J}_{n m_n,\, jm_j}\,
		\Cc^{J\lambda}_{n\lambda,\,j0} \, .
		\label{eq:angular_integral_result}
\end{align}
The angular integral over $\bh{\p}$ gives a similar result. 
Therefore, combining these results with the sums over all $m_J$, $m_j$, $m_{\ell'}$, $m_{\ell}$ in Eq.~\eqref{eq:H_spin_helicity}, we find
\begin{align}
	\sum_{m_J,m_j}
	\sum_{m_{\ell'},m_{\ell}} 
	&\left(\Ic^{J j \ell\ell'}_{m_J m_j m_{\ell} m_{\ell'}}(\lambda')\right)^*
	\Ic^{J j \ell'\ell}_{m_J m_j m_{\ell'} m_{\ell} }(\lambda) 
	\nn \\[5pt]
	& = 
	\frac{1}{(2J+1)^2}
	\sum_{n',n}
	\Cc^{n'\lambda'}_{\ell'\lambda',\,\ell0} \,
	\Cc^{J\lambda'}_{n'\lambda',\,j0} \,
	\Cc^{n\lambda}_{\ell\lambda,\,\ell'0} \,
	\Cc^{J\lambda}_{n\lambda,\,j0} \,
	\nn \\[5pt]
	& \qquad \times 
	\sum_{m_{n'},m_n}
	\sum_{m_j,m_J}
	\Cc^{Jm_J}_{n'm_{n'},\,jm_j}\,
	\Cc^{Jm_J}_{n m_n,\,jm_j} \,
	\sum_{m_{\ell'},m_{\ell}}
	\Cc^{n'm_{n'}}_{\ell'm_{\ell'},\,\ell m_{\ell}}\,
	\Cc^{n m_n}_{\ell m_{\ell},\,\ell'm_{\ell'}}\,
	.
\end{align}
Again, since all quantum numbers are integers, the square of the phase $(-1)^{\ell'+\ell+j-J}$ appearing in Eq.~\eqref{eq:angular_integral_result} is unity. 
The last line of sums over Clebsch-Gordan coefficients are reduced via iterations of the orthonormality of the coefficients, \cf~Eq.~\eqref{eq:app_cg_orthogonality_1}. 
Imposing these sequential orthogonality conditions, as well as the symmetry condition Eq.~\eqref{eq:app_cg_sym_2} and the permutation identity Eq.~\eqref{eq:app_cg_zero_projection_permutation} to order the quantum number arguments in a convenient form, the sum over angular integrals reduces to 
\begin{align}
	\sum_{m_J,m_j}
	\sum_{m_{\ell'},m_{\ell}} 
	&\left(\Ic^{J j \ell\ell'}_{m_J m_j m_{\ell} m_{\ell'}}(\lambda')\right)^*
	\Ic^{J j \ell'\ell}_{m_J m_j m_{\ell'} m_{\ell} }(\lambda) 
	\nn \\[5pt]
	& =
	\sum_{n = \lvert \ell' - \ell \rvert}^{\ell'+\ell}
	\frac{(-1)^{\ell'+\ell-n}}{2n+1}\,
	\Cc_{J\lambda,\,j0}^{n\lambda} \,
	\Cc_{\ell\lambda,\,\ell'0}^{n\lambda} \,
	\Cc_{J\lambda',\,j0}^{n\lambda'} \, 
	\Cc_{\ell'\lambda',\,\ell0}^{n\lambda'} \, .
\end{align}
The bounds on the sum are due to the angular momentum selection rules, $\lvert\ell'-\ell\rvert\le n \le \ell'+\ell$.

We therefore conclude that the $\Hc_{S'\lambda',\,S\lambda}^{j,J}$ is given by the expression
\begin{align}
	\Hc^{j,J}_{S'\lambda',\,S\lambda}(p,k) 
	& = 
	\sum_{\ell'=0}^{S'} 
	\sum_{\ell=0}^{S} 
	\sum_{n = \lvert\ell'-\ell\rvert}^{\ell'+\ell}\,
	\frac{(-1)^{n}}{2n+1}\,
	\left(\frac{k}{q_p^\star}\right)^{S'}
	\sqrt{2\ell'+1} \,
	\Bc_{S'\lambda'}^{\ell'}(k;p) \, 
	\Cc_{J\lambda',\,j0}^{n\lambda'} \, 
	\Cc_{\ell'\lambda',\,\ell0}^{n\lambda'} \,
	\nn \\[5pt]
	& \qquad \times
	\left(\frac{p}{q_k^\star}\right)^{S}
	\sqrt{2\ell+1} \,
	\Bc_{S\lambda}^{\ell}(p;k) \,
	\Cc_{J\lambda,\,j0}^{n\lambda} \,
	\Cc_{\ell\lambda,\,\ell'0}^{n\lambda} \, .
    \label{eq:H_helicity_final}
\end{align}
The angular-momentum selection rules further require $\lvert n-J\rvert\leq j\leq n+J$. 
Since $\ell\leq S$ and $\ell'\leq S'$, the largest possible value of $j$ is $J+S+S'\equiv \jmaxcg$. 
Thus $\Hc=0$ for $j>\jmaxcg$, and the Neumann series in Eq.~\eqref{eq:ope_pw_helicity} terminates. 
Transforming to the spin-orbit basis with Eq.~\eqref{eq:ope_hel_to_pw} then gives the main result, Eq.~\eqref{eq:ope_result}, with $\Hc^{j,J}_{L'S',LS}$ given by Eq.~\eqref{eq:ope_HLS}.

Reference~\cite{Jackura:2023qtp} presented the partial-wave OPE in the form
\begin{align}
    \Gc^{J^P}_{L'S',\, LS}(p,k) 
    = 
    \Kc^{J^P}_{\Gc;L'S',\, LS}(p,k) 
    + 
    \Tc^{J^P}_{L'S',\, LS}(p,k) \, 
    Q_0(\zeta_{pk})
    \label{eq:ope_split_form}
\end{align}
where $\Kc_{\Gc}$ and $\Tc$ are regular functions of $p$ and $k$. 
Equation~\eqref{eq:ope_result} can be connected back to the main result of the prior OPE study using recursive properties of the $Q_j$ functions,
\begin{align}
    Q_j(z) 
    = 
    P_j(z) Q_0(z) 
    - 
    \sum_{n = 1}^j 
    \frac{1}{n}P_{n-1}(z) P_{j-n}(z) \, ,
    \label{eq:bonnet}
\end{align}
where $P_j$ are Legendre functions of the 1st kind and $Q_0$ is explicitly
\begin{align}
	Q_0(z) = \frac{1}{2} \log\left(\frac{z+1}{z-1}\right) \, .
\end{align}
Applying this expression to Eq.~\eqref{eq:ope_result} yields the following for the $\Kc_{\Gc}^{J^P}$ and $\Tc^{J^P}$ coefficients of Ref.~\cite{Jackura:2023qtp},
\begin{subequations}
\begin{align}
    \Kc_{\Gc; L'S',\, LS}^{J^P}(p,k) 
    &= 
    -\frac{1}{2pk} 
    \sum_{j=0}^{\jmaxcg} (2j+1)\, 
    \Hc^{j,J}_{L'S',\, LS}(p,k) 
    \sum_{n = 1}^j 
    \frac{1}{n}
    P_{n-1}(\zeta_{pk}) \, 
    P_{j-n}(\zeta_{pk}) \, , \\
    \Tc_{L'S',\, LS}^{J^P}(p,k) 
    &= 
    \frac{1}{2pk} 
    \sum_{j=0}^{\jmaxcg} (2j+1)\, 
    \Hc^{j,J}_{L'S',\, LS}(p,k) \, 
    P_j(\zeta_{pk}) \, .
\end{align}
\end{subequations}
%

\subsection{Boost Coefficients}
\label{sec:boost_coefficients}

The remaining ingredient is an explicit expression for the boost coefficients introduced in Eq.~\eqref{eq:solid_harmonic_expansion}. Inverting that expansion gives
\begin{align}
	\Bc_{S\lambda}^\ell (p;k) 
    = 
    \int \diff \bh{\p} \, 
    Y^*_{\ell \lambda}(\bh{\p}) \,
    Y_{S{\lambda}}(\hatpstark) \, 
    \left(\frac{\pstark}{p}\right)^S \, . 
    \label{eq:boost_coefficient_def}
\end{align}
where $\p_k^\star=\p_k^\star(\p)$ is obtained from the Lorentz boost in Eqs.~\eqref{eq:recap_boost_par} and~\eqref{eq:recap_boost_perp}. 
Both spherical harmonics are evaluated with respect to the helicity axis $-\bh{\k}$. 
Note that when $\bs{\beta}_k=\0$, one has $\p_k^\star=\p$, and Eq.~\eqref{eq:boost_coefficient_def} gives $\Bc^\ell_{S\lambda}=\delta_{\ell S}$, as expected. 
Throughout this section, all momenta have their angles defined with respect to the helicity quantization axis, thus we drop the prime ($'$) notation introduced in Eq.~\eqref{eq:solid_harmonic_expansion} for notational convenience.

We construct the recursive definition using the Clebsch-Gordan series, \cf~App.~\ref{app:sec:ang_mom}. For $S=0$, the angular distribution is constant, $Y_{00}=1/\sqrt{4\pi}$, and hence $\Bc^\ell_{0\lambda}(p;k)=\delta_{\ell0}\delta_{\lambda0}$. 
The first nontrivial coefficient, and the base case for the recursion, occurs at $S=1$. Using
\begin{align}
    \label{eq:boosted_p_spherical_basis}
	p_k^\star \,Y_{1\lambda}(\hatpstark) = \sqrt{\frac{3}{4\pi}} \, (\p_k^\star)_\lambda \, ,
\end{align}
with $(\p_k^\star)_\lambda$ the spherical component in the basis whose longitudinal axis is $\hat{\bs{\beta}}_k=-\bh{\k}$.~\footnote{The spherical components of a vector $\p$ in terms of their cartesian components are
\begin{align}
	\p_{\pm 1} = \mp \frac{1}{\sqrt{2}}(\p_x \pm i\p_y)\, , 
	\qquad 
	\p_0 = \p_z \, . \nn 
\end{align}
}
The transverse components are unchanged by the boost,
\begin{align}
    \label{eq:boosted_sph_harm_pm_one}
	p_k^\star \,Y_{1,\pm 1}(\hatpstark) = p Y_{1,\pm 1}(\bh{\p}) \, .
\end{align}
The longitudinal component, however, is modified according to Eq.~\eqref{eq:recap_boost_par}, giving
\begin{align}
    \label{eq:boosted_sph_harm_zero}
	p_k^\star \,Y_{1 0}(\hatpstark) = \gamma_k p\, Y_{10}(\bh{\p}) - \sqrt{3}\, \gamma_k \beta_k \omega_p \,  Y_{00}(\bh{\p}) \, .
\end{align}
As expected, the expansion does not mix the helicity quantum number, and it involves angular functions $\ell \le S = 1$. 
Therefore, we find explicitly that the expansion is
\begin{align}
	p_k^\star \,Y_{1\lambda}(\hatpstark) = p\sum_{\ell = 0}^{1} \Bc^{\ell}_{1\lambda}(p;k)\, Y_{\ell \lambda}(\bh{\p}) \, ,
\end{align}
with boost coefficients
\begin{align}
	\Bc_{1 \lambda}^{\ell}(p;k) 
	= \delta_{\ell 1} 
	\left( \delta_{\lambda,1} + \delta_{\lambda,-1} + \gamma_k \delta_{\lambda 0} \right)  
	- 
	\sqrt{3}\,
	\frac{\gamma_k \beta_k \omega_p}{p} \, 
	\delta_{\ell 0}\delta_{\lambda 0} \, .
\end{align}

Having established the $S=1$ case, we derive the recursion for general $S$. Starting from Eq.~\eqref{eq:boost_coefficient_def}, we decompose $Y_{S\lambda}(\hatpstark)$ into spherical harmonics of ranks $S-1$ and $1$ using Eq.~\eqref{eq:app_cg_ylm_decomp}. 
This gives
\begin{align}
    p^S \Bc_{S\lambda}^\ell (p; k) 
    &= 
    \frac{1}{\Cc^{S 0}_{S-1,0,\, 10}} \, 
    \sqrt{\frac{4\pi(2S+1)}{3(2S-1)}} \,
    \sum_{\mu_a=-(S-1)}^{S-1} \sum_{\mu_b = -1}^1 \Cc^{S \lambda}_{S-1, \mu_a,\, 1\mu_b}
    \nn\\[5pt]
    &\qquad\qquad\times \int \! \diff \bh{\p} \, 
    Y^*_{\ell \lambda}(\bh{\p})\, 
    \big[
    (\pstark)^{S-1}\, 
    Y_{S-1, \mu_a}(\hatpstark)
    \big]
    \big[
    \pstark\,  
    Y_{1{\mu_b}}(\hatpstark)
    \big] \,
     \, .
\end{align}
Now, we apply Eq.~\eqref{eq:solid_harmonic_expansion} to the rank-1 and rank-$(S-1)$ solid harmonics in the square brackets which are summed. 
We therefore find 
\begin{align}
  \Bc_{S\lambda}^\ell (p; k) &= \sqrt{\frac{4\pi(2S+1)}{3S}}\, 
  \sum_{\mu_a,\mu_b} 
  \Cc^{S\lambda}_{S-1, \mu_a,\, 1\mu_b} 
  \sum_{a=0}^{S-1} \sum_{b=0}^1 
  \Bc^{a}_{S-1,\, \mu_a}(p; k) \, 
  \Bc^{b}_{1 \mu_b}(p;k) 
  \nn \\[5pt]
  &\qquad \times 
  \int \!\diff \bh{\p} \, 
  Y^*_{\ell \lambda}(\bh{\p}) \, 
  Y_{a \mu_a}(\bh{\p}) \, 
  Y_{b \mu_b}(\bh{\p}) \,,
  \label{eq:boost_recursive_gaunt}
\end{align}
where we used $\Cc^{S0}_{S-1,0,\,10}=\sqrt{S/(2S-1)}$. 
Evaluating the remaining angular integral as the Gaunt coefficient in Eq.~\eqref{eq:app_cg_gaunt}, Eq.~\eqref{eq:boost_recursive_gaunt} reduces to the recurrence relation in Eq.~\eqref{eq:boost_coeff_recursive}.

The recursion may also be iterated into an explicit product form. 
Let $r$ denote the rank after the $r$th Clebsch-Gordan decomposition, with intermediate labels $\ell_r$ and $\Lambda_r$. 
Each step couples the rank-$(r-1)$ result to one rank-one boosted harmonic,
\begin{align}
    \Bc^{\ell_r}_{r\Lambda_r}(p;k)
    &=
    \sum_{\ell_{r-1}=0}^{r-1}
    \sum_{b_r=0}^{1}
    \sum_{\Lambda_{r-1},\mu_r}
    \sqrt{
    \frac{(2r+1)(2\ell_{r-1}+1)(2b_r+1)}
    {3r(2\ell_r+1)}
    }
    \nn\\
    &\qquad\qquad\times
    \Cc^{r\Lambda_r}_{r-1,\Lambda_{r-1},\,1\mu_r} \, 
    \Cc^{\ell_r0}_{\ell_{r-1}0,\,b_r0} \,
    \Cc^{\ell_r\Lambda_r}_{\ell_{r-1}\Lambda_{r-1},\,b_r\mu_r} \, 
    \Bc^{\ell_{r-1}}_{r-1,\Lambda_{r-1}}(p;k)
    \Bc^{b_r}_{1\mu_r}(p;k) \, . \nn
\end{align}
Here $b_r=0,1$, $\mu_r=-1,0,1$, and the remaining sums have their usual angular-momentum ranges. 
The initial step is fixed by $\ell_0=0$, $\Lambda_0=0$, and $\Bc^0_{00}=1$. 
The Clebsch-Gordan coefficients enforce $\Lambda_r=\Lambda_{r-1}+\mu_r$ and the usual triangle conditions. 
Iterating this expression to $r=S$, with $\ell_S=\ell$, gives
\begin{align}
    \Bc^\ell_{S\lambda}(p;k)
    &=
    \sum_{\ell_1=0}^{1}\cdots\sum_{\ell_{S-1}=0}^{S-1}
    \sum_{b_1,\ldots,b_S=0}^{1}
    \sum_{\mu_1,\ldots,\mu_S=-1}^{1}
    \delta_{\lambda,\Lambda_S}
    \prod_{r=1}^{S}
    \sqrt{
    \frac{(2r+1)(2\ell_{r-1}+1)(2b_r+1)}
    {3r(2\ell_r+1)}
    }
    \nonumber\\
    &\qquad\qquad\times
    \Cc^{r\Lambda_r}_{r-1,\Lambda_{r-1},\,1\mu_r} \, 
    \Cc^{\ell_r0}_{\ell_{r-1}0,\,b_r0} \,
    \Cc^{\ell_r\Lambda_r}_{\ell_{r-1}\Lambda_{r-1},\,b_r\mu_r} \, 
    \Bc^{b_r}_{1\mu_r}(p;k)
    \,,
    \label{eq:boost_coeff_product}
\end{align}
where $\Lambda_r=\sum_{i=1}^r \mu_i$. 
Thus, the full boost coefficient is a finite sum of products of the elementary $S=1$ boost coefficients, with the Clebsch-Gordan chain tracking how the rank-one factors build the final $(S,\lambda)$ and $(\ell,\lambda)$ quantum numbers.

\subsection{Threshold behavior}
\label{sec:threshold}

We now check that our result has the expected behavior in the relevant threshold limits. 
There are two distinct cases. 
First, when $p$ and $k$ are fixed and the pair invariant masses approach their two-body thresholds, the behavior is governed by the internal pair angular momenta $S$ and $S'$. 
Second, when $\sigma_p$ and $\sigma_k$ are held fixed above their respective thresholds and $p,k\to0$, the behavior is governed by the orbital angular momenta $L$ and $L'$.

To start, consider a generic $\3\to\3$ amplitude $\Mc_{3;L'S',\,LS}^{J^P}(p,k)$ with $p$ and $k$ fixed above zero, with the pair invariant masses $\sigma_p$ and $\sigma_k$ approaching their respective thresholds. 
Because the two-body subsystems are associated with spherical harmonics in the angles of the momenta $\q_k^\star$ and $\q_p^\star$, the amplitude must behave near threshold as $\Mc_{3;L'S',\,LS}^{J^P}(p,k) \sim (q_p^\star)^{S'}(q_k^\star)^{S}$ as $q_k^\star,q_p^\star \to 0$ for fixed $p$ and $k$~\cite{Briceno:2024ehy}. 
The isolated OPE kernel $\Gc^{J^P}_{L'S',\,LS}$ has a different behavior, as the OPE contribution to the full amplitude has the functional form $\Mc_{2;S'} \, \Gc^{J^P}_{L'S',\,LS} \, \Mc_{2;S}$ for some $\2\to\2$ partial-wave amplitude $\Mc_{2,S}$, \cf~Ref.~\cite{Jackura:2022gib,Jackura:2023qtp}. 
Since each of the two-body partial-wave amplitudes have the threshold behavior $\Mc_{2,S}(\sigma_k) \sim (q_k^\star)^{2S}$ as $q_k^\star \to 0$, the OPE must behave as $\Gc^{J^P}_{L'S',\,LS}(p,k) \sim (q_p^\star)^{-S'}(q_k^\star)^{-S}$ as $q_k^\star,q_p^\star \to 0$ for fixed $k$ and $p$. 
This behavior is the origin of the barrier factors in Eq.~\eqref{eq:recap_ope_def}, thus by definition the OPE respects this threshold behavior, see Refs.~\cite{Hansen:2014eka,Jackura:2022gib}.

We now consider the threshold behavior of the $\3\to\3$ amplitude for $\sigma_p$ and $\sigma_k$ held fixed away from the two-body thresholds, so that $\qstarp$ and $\qstark$ are nonzero constants. 
Then, the $\3\to\3$ partial-wave amplitude is effectively a quasi-two-body process consisting of the spectators and some quasiparticles of fixed masses $\sqrt{\sigma_k}$ and $\sqrt{\sigma_p}$. 
As the system approaches the quasi-two-body threshold, that is $p,k\to 0$, the partial-wave amplitude must have at least the familiar leading order momentum scaling $\Mc_{3;L'S',\,LS}^{J^P}(p,k) = \Oc(p^{L'}k^L)$ as $p,k \to 0$ for fixed pair invariant masses above thresholds. 
The OPE behaves in a similar way, 
\begin{align}
	\Gc_{L'S',\,LS}^{J^P}(p,k) 
	= 
	\Oc(p^{L'}k^L) \,\, \textrm{as } p,k\to 0,
	\label{eq:ope_threshold_expected}
\end{align}
for fixed pair invariant masses above thresholds. 
We note that the OPE for some target $J^P$ could have a threshold behavior more suppressed than this minimal requirement.

We now verify that the finite-sum representation in Eq.~\eqref{eq:ope_result} has the same behavior as Eq.~\eqref{eq:ope_threshold_expected}. 
To capture the correct power-counting, it is useful to organize the simultaneous small $p,k$ expansion by total degree. 
We write $\Oc(N)$ for terms of combined degree $N$ or higher in $p$ and $k$, with $\sigma_p$ and $\sigma_k$ held fixed. 
For example, if a correction is $\Oc(p^3k^2)$, then we write $\Oc(5)$.

For small $p$ and $k$, the argument $\zeta_{pk}$ defined in Eq.~\eqref{eq:recap_zeta} behaves like $\zeta_{pk} = \Delta/(2pk) + \Oc(0)$ with $\Delta \equiv u_{pk}\rvert_{p=k=0} - m_e^2$, assuming that we are away from the on-shell point $\Delta = 0$. 
Thus, as $p,k\to 0$, $\zeta_{pk} \to \infty$. Using the asymptotic expansion of $Q_j(z)$ for $z\to \infty$,
\begin{align}
	Q_j(z) = \frac{2^j (j!)^2}{(2j+1)!} \, z^{-j-1} + \Oc(z^{-j-3})\, ,
\end{align}
we find that
\begin{align}
	\frac{2j+1}{2pk}\,
	Q_j(\zeta_{pk})
	&=
	\alpha_j \,
	p^j k^j
	+
	\Oc(2j+2) \, ,
	\label{eq:Qterm_threshold}
\end{align}
as $p,k\to 0$, with $\alpha_j$ absorbing all the non-momentum dependence, $\alpha_j \equiv 2^{2j} (j!)^2 \Delta^{-j-1} / (2j)! $.

For the $j$th term in the sum of Eq.~\eqref{eq:ope_result}, we expect that $\Hc^{j,J^P}_{L'S',\,LS}$ must contribute the remaining powers. 
This can be seen directly by first examining the threshold form of the boost coefficients. 
It is easiest to do so with the integral representation Eq.~\eqref{eq:boost_coefficient_def}, from which we investigate how $\Yc_{S\lambda}(\p_k^\star)$ behaves for small $p$ and $k$. In the helicity frame associated with $-\bh{\k}$, the boost to the initial-pair rest frame is, to leading order, a translation along the quantization axis. 
To see this, we combine Eqs.~\eqref{eq:recap_boost_par} and~\eqref{eq:recap_boost_perp} into 
\begin{align}
	\p_k^\star
	&=
	\p+
	\left[
	\frac{\gamma_k-1}{\beta_k^2}
	(\p\cdot\bs{\beta}_k)
	-\gamma_k\,\omega_p
	\right]\bs{\beta}_k \, ,
	\label{eq:lorentz_boost}
\end{align}
where $\bs{\beta}_k= -\k / (\sqrt{s} - \omega_k)$ and $\gamma_k = 1/\sqrt{1-\beta_k^2}$. For small $\k$, $\beta_k = \Oc(k)$, and $(\gamma_k-1)/\beta_k^2$ has a regular Taylor expansion about $\beta_k=0$, while $\omega_p = m_p + \Oc(p^2)$ for small $\p$. Thus, $\p_k^\star$ is regular in both $\p$ and $\k$ for small momenta, with the same reasoning applying to $\k_p^\star$.
A simultaneous expansion in both small $\p$ and $\k$ thus gives the leading-order behavior 
\begin{align}
	\p_k^\star = \p + \frac{m_p}{\sqrt{\sigma_k}}\,\k + \Oc(3) \, .
	\label{eq:lorentz_boost_threshold}
\end{align}
By inspection of Eq.~\eqref{eq:lorentz_boost}, the $\Oc(3)$ term is proportional to $\k$. 

Since the regular solid harmonics are homogeneous polynomials of their arguments~\cite{STEINBORN19731}, the numerator of Eq.~\eqref{eq:recap_ope_def} is then analytic in $\p$ and $\k$. 
Therefore, inserting Eq.~\eqref{eq:lorentz_boost_threshold} into the solid harmonic and Taylor expanding yields the leading order behavior
\begin{align}
	\Yc_{S\lambda}(\p_k^\star) 
	& = 
	\Yc_{S\lambda}\left(\p+(m_p/\sqrt{\sigma_k})\, \k \right) + \Oc(S+2)\, ,
	\label{eq:leading_solid_harmonic}
\end{align}
where the first term is of degree $S$ while the derivative term in the Taylor expansion has degree $S-1$, giving the total correction $3+(S-1) = S+2$.~\footnote{Explicitly, the Taylor expansion for the solid harmonic is
\begin{align}
	\Yc_{\ell m}(\x + \delta \x) = \Yc_{\ell m}(\x) + \delta\x \cdot \bs{\nabla}_\y\Yc_{\ell m}(\y) \,\Big\rvert_{\y = \x} + \Oc(\delta x^2) \, .
\end{align}
}
Now, the argument of the leading term in Eq.~\eqref{eq:leading_solid_harmonic} is a linear translation, thus we can use the addition theorem for solid harmonics, Eq.~\eqref{eq:app_cg_solid_harmonic_addition}, to separate the $\p$ and $\k$ dependence~\cite{STEINBORN19731},
\begin{align}
	\Yc_{S\lambda}\left(\p+(m_p/\sqrt{\sigma_k})\, \k \right) 
	& =
	\sum_{\ell = 0}^{S} 
	\sqrt{\frac{4\pi}{2\ell+1}\,\binom{2S+1}{2\ell}}\,
	\nn \\[5pt]
	& \qquad  \times
	\sum_{\mu = -\ell}^{\ell}
	\Cc^{S\lambda}_{\ell\mu,\, S-\ell,\lambda-\mu}\,
	\Yc_{\ell\mu}(\p) \,
	\Yc_{S-\ell,\,\lambda-\mu}((m_p/\sqrt{\sigma_k})\,\k) \, .
	\label{eq:solid_harmonic_addition}
\end{align}
We now use the fact that since the linear shift is itself along the quantization axis $-\bh{\k}$, the solid harmonic of this shift simplifies to 
\begin{align}
	\Yc_{S-\ell,\lambda-\mu}((m_p /\sqrt{\sigma_k})\,\k) 
	=
	(-1)^{S-\ell}\,
	\delta_{\lambda-\mu, 0} \, 
	\left(\frac{m_p\,k}{\sqrt{\sigma_k}}\right)^{S-\ell}\,
	\sqrt{\frac{2(S-\ell)+1}{4\pi}} \, ,
	\label{eq:solid_harmonic_simplify_z}
\end{align}
where the $(-1)^{S-\ell}$ phase is because the helicity axis for the pair is $-\bh{\k}$ while the leading shift in Eq.~\eqref{eq:lorentz_boost_threshold} is proportional to $+\k$.
Combining Eqs.~\eqref{eq:solid_harmonic_addition} and~\eqref{eq:solid_harmonic_simplify_z} with the leading order expression, Eq.~\eqref{eq:leading_solid_harmonic} gives
\begin{align}
	\Yc_{S\lambda}(\p_k^\star)
	& = 
	\sum_{\ell = 0}^{S} 
	(-1)^{S-\ell}\,
	\sqrt{\binom{2S+1}{2\ell + 1} } \,
	\Cc^{S\lambda}_{\ell \lambda,\,S-\ell, 0} \,
	\left(\frac{m_p\,k}{\sqrt{\sigma_k}}\right)^{S-\ell} \,
	\Yc_{\ell \lambda}(\p) \, 
	+
	\Oc(S+2) \, ,
	\label{eq:boosted_solid_harmonic_leading_order}
\end{align}
for $p$ and $k$ approaching zero.

Equation~\eqref{eq:boosted_solid_harmonic_leading_order} can be used directly in the definition of the boost coefficient, Eq.~\eqref{eq:boost_coefficient_def}, which upon using the orthogonality of the spherical harmonics, Eq.~\eqref{eq:app_cg_ylm_orthogonality}, results in leading order behavior for the boost coefficients
\begin{align}
	p^S\Bc_{S\lambda}^{\ell} (p;k) 
    = 
    (-1)^{S-\ell}\,
	\sqrt{\binom{2S+1}{2\ell + 1} } \,
	\left(\frac{m_p}{\sqrt{\sigma_k}}\right)^{S-\ell} \,
	\Cc^{S\lambda}_{\ell \lambda,\,S-\ell, 0} \,
	p^{\ell} k^{S-\ell}
	+
	\Oc(S+2)
    \, ,
    \label{eq:boost_coeff_threshold}
\end{align}
as $p,k\to 0$. 
We note that it is necessary to keep track of the Clebsch-Gordan coefficient as it provides necessary restrictions on the quantum numbers appearing in various sums. 
Notice that when $k=0$ for fixed $p$, we find that the boost coefficient is zero unless $\ell = S$. In this case, Eq.~\eqref{eq:boost_coeff_threshold} becomes the identity, consistent with the exact result from Eq.~\eqref{eq:boost_coefficient_def} when $\k = \0$ for any finite $\p$.

Using Eq.~\eqref{eq:boost_coeff_threshold} in the vertex factor defined in Eq.~\eqref{eq:ope_V}, the initial-state vertex has the threshold scaling
\begin{align}
    \Vc^{n,j,J}_{LS}(p,\ell;k,\ell')
    &=
    p^\ell k^{S-\ell} \,
    A^{n,J}_{LS,\,\ell}(\sigma_k) \,
    \Sigma^{n,j,J}_{LS}(\ell;\ell') \,
    + 
    \Oc(S+2) 
    \, ,
    \label{eq:threshold_red_vertex_initial_coeff}
\end{align}
where $A^{n,J}_{LS,\,\ell}$ is the momentum-independent coefficient
\begin{align}
	A^{n,J}_{LS,\,\ell}(\sigma_k) 
	= 
	(-1)^{S-\ell}\,
	\sqrt{\frac{(2L+1)(2S+1)}{(2J+1)(2n+1)} \binom{2S}{2\ell}}\,
    \left(\frac{1}{q_k^\star}\right)^{S} 
	\left(\frac{m_p}{\sqrt{\sigma_k}}\right)^{S-\ell} \, ,
\end{align}
while $\Sigma^{n,j,J}_{LS}$ is the helicity sum
\begin{align}
    \Sigma^{n,j,J}_{LS}(\ell;\ell')
	= 
	\sum_{\lambda} 
    \,
    \Cc^{S\lambda}_{\ell \lambda,\,S-\ell, 0} \,
    \Cc^{J\lambda}_{L0,\,S\lambda} \,
    \Cc^{n\lambda}_{J\lambda,\,j0} \,
    \Cc^{n\lambda}_{\ell\lambda,\,\ell'0} \, .
    \label{eq:helicity_sum}
\end{align}
With an appropriate change of variables, the final-state vertex function has a similar behavior
\begin{align}
    \Vc^{n,j,J}_{L'S'}(k,\ell';p,\ell)
    &=
    k^{\ell'} p^{S'-\ell'} \,
    A^{n,J}_{L'S',\,\ell'}(\sigma_p) \,
    \Sigma^{n,j,J}_{L'S'}(\ell';\ell)
    + 
    \Oc(S'+2) \,.
    \label{eq:threshold_red_vertex_final_coeff}
\end{align}
Inserting Eqs.~\eqref{eq:threshold_red_vertex_initial_coeff} and~\eqref{eq:threshold_red_vertex_final_coeff} into Eq.~\eqref{eq:ope_HLS}, we find the leading behavior for $\Hc^{j,J^P}$,
\begin{align}
	\Hc_{L'S',\,LS}^{j,J^P}(p,k)
	& =
	\sum_{\ell',\ell} \,
	p^{S'+\ell - \ell'} k^{S+\ell'-\ell}
    \sum_{n}
	(-1)^{n} \,
	\nn \\[5pt]
	& \quad \times 
	A^{n,J}_{L'S',\,\ell'}(\sigma_p) \,
	A^{n,J}_{LS,\,\ell}(\sigma_k) \,
    \Sigma^{n,j,J}_{L'S'}(\ell';\ell) \,
    \Sigma^{n,j,J}_{LS}(\ell;\ell')
	+
	\Oc(S' + S + 2) \,.
	\label{eq:H_threshold}
\end{align}

The helicity sums in Eq.~\eqref{eq:H_threshold} may be expressed in a series of Racah-recouplings such that the rank-$\ell$ and rank-($S'-\ell'$) tensors first couple to some intermediate rank, which then couples with the exchange rank-$j$ to the final spectator orbital rank-$L'$. 
The rank-$\ell'$ and rank-($S-\ell$) tensors couple similarly to orbital rank $L$. 
We demonstrate these Racah-recouplings explicitly in App.~\ref{app:sec:helicity_recoupling} for the interested reader.
Those recouplings establish Clebsch–Gordan triangle conditions that imply
\begin{align}
	S' + \ell - \ell' \ge \lvert L' - j\rvert \, , \textrm{ and } S + \ell' - \ell \ge \lvert L - j\rvert \, .
\end{align}
We now impose the relations $\lvert L-j\rvert \ge \max(L-j,0)$ and $\lvert L'-j\rvert \ge \max(L'-j,0)$, which imply the weaker, but sufficient for our purposes, power-counting bounds,
\begin{align}
	S' + \ell - \ell' \ge \max(L'-j,0) \, , \textrm{ and } S + \ell' - \ell \ge \max(L-j,0) \, .
\end{align}
If either condition fails, no admissible intermediate angular momentum exists, and the corresponding low-power term vanishes identically. 

These bounds imply that every nonvanishing term in Eq.~\eqref{eq:H_threshold}
contains at least $\max(L'-j,0)$ powers of $p$ and at least
$\max(L-j,0)$ powers of $k$. Hence
\begin{align}
	\Hc^{j,J^P}_{L'S',\,LS}(p,k)
	=
	\Oc\!\left(
	p^{\max(L'-j,0)}k^{\max(L-j,0)}
	\right) \,.
\end{align}
Combining this with Eq.~\eqref{eq:Qterm_threshold}, the $j$th term scales as 
\begin{align}
	\Gc_{L'S',\,LS}^{j,J^P}(p,k)
	=
	\Oc\!\left(
	p^{\max(j,L')}k^{\max(j,L)}
	\right) \, .
    \label{eq:ope_threshold_final}
\end{align}
where we used $j+\max(L'-j,0)=\max(L',j)$ and $j+\max(L-j,0)=\max(L,j)$. Since $\max(j,L')\geq L'$ and $\max(j,L)\geq L$, every term in the finite sum is at least of order $p^{L'}k^L$. Individual terms can be further suppressed by angular-momentum selection rules or by cancellations among the spin-orbit couplings in Eq.~\eqref{eq:H_threshold}. The finite sum therefore has the expected behavior $\Gc^{J^P}_{L'S',\,LS}=\Oc(p^{L'}k^L)$ as $p,k\to0$, in agreement with Eq.~\eqref{eq:ope_threshold_expected} and with Ref.~\cite{Jackura:2023qtp}.

\section{Numerical Studies}
\label{sec:numerical}

We conclude with numerical studies of Eq.~\eqref{eq:ope_result} for the OPE function and Eq.~\eqref{eq:boost_coeff_recursive} for the boost coefficients. 
These examples verify the expected kinematic behavior of the formulae and illustrate their applicability to higher-angular-momentum channels relevant for three-pion systems. 
We work in the isospin-symmetric limit, taking all pion masses to be degenerate, $m_\pi\equiv m$. 
The kinematics therefore reduce to $m_p=m_k=m_e\equiv m$, and since pions are pseudoscalars, the intrinsic-parity factor is $\eta=-1$. 
The elastic region is then bounded by the three- and five-pion thresholds, $(3m)^2 \leq s < (5m)^2$.

The quantities plotted below are the flavor-stripped OPE functions. 
In a realistic $3\pi$ calculation, the spin-orbit amplitudes must be embedded in channel states with definite isospin and symmetrized according to Bose symmetry. 
These symmetry restrictions determine which combinations of pair isospin, pair angular momentum, and spectator-pair orbital angular momentum are allowed. 
The corresponding isospin recoupling factor multiplies the OPE function, as discussed in Sec.~VI of Ref.~\cite{Jackura:2023qtp}. 
We do not include this factor in the plots for simplicity. 
We do, however, impose the appropriate three-pion symmetry restrictions on the displayed quantum numbers and focus on the universal kinematic and angular-momentum structure of the OPE.
\begin{figure}
    \centering
    \includegraphics[width=0.45\textwidth]{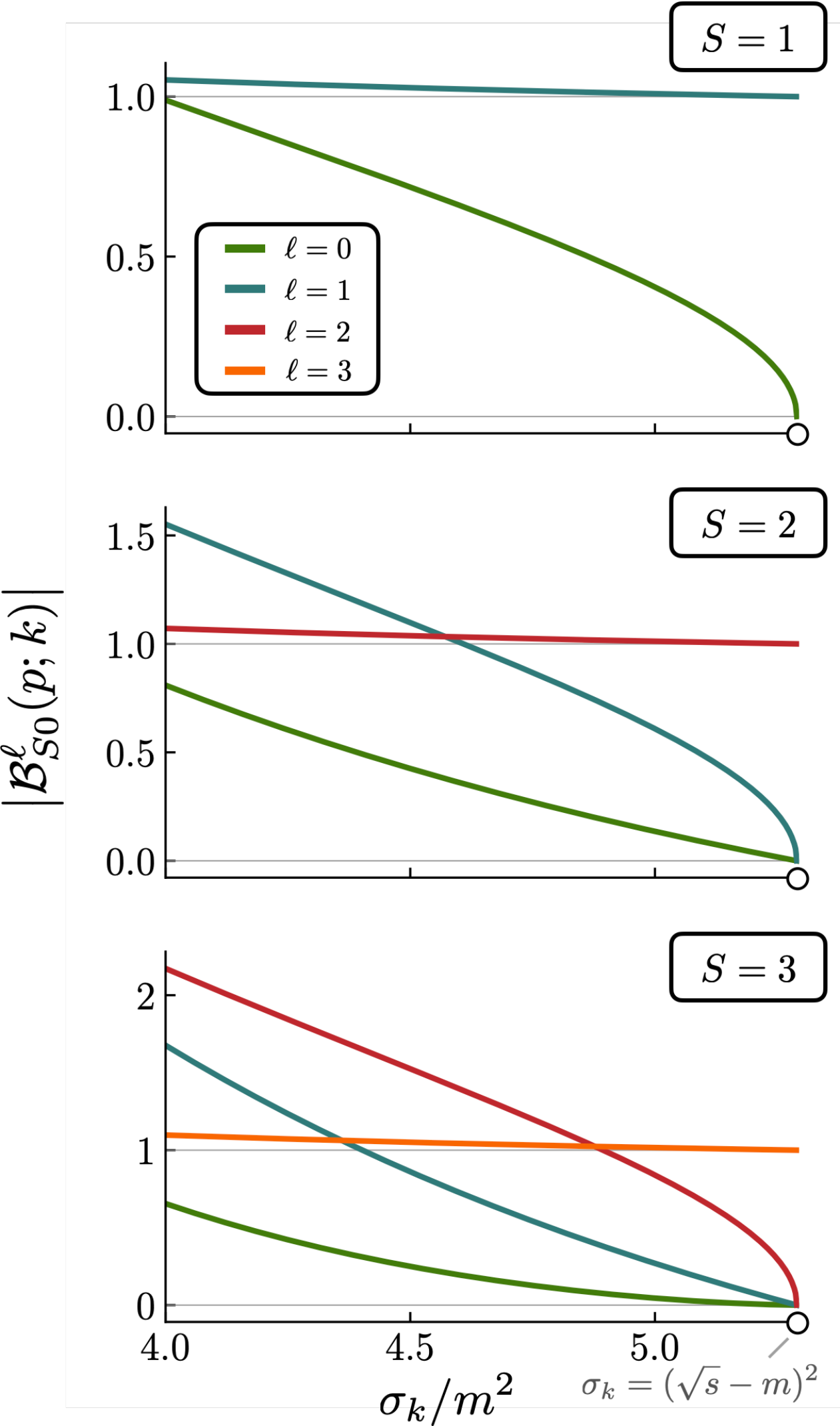}
    \caption{Absolute value of boost coefficients $\Bc^\ell_{S \lambda}(p;k)$ for $\lambda = 0$, $\sqrt s / m = 3.3$, $p/m = 0.7$, and $4 \leq \sigma_k / m^2 \leq (\sqrt s / m - 1)^2 = 5.29$. 
    Each panel corresponds to a different choice of $S$, from $S = 1$ to $S = 3$, and within each panel the coefficient for each value of $\ell \leq S$ is included. 
    The upper bound of the physical region in $\sigma_k$ is marked.}
    \label{fig:boost_coeffs}
\end{figure}

\subsection{Numerical verification of threshold behaviors}
\label{sec:numerical_verification_threshold}

We first examine the threshold behavior of the boost coefficients $\Bc_{S\lambda}^{\ell}$ defined in Eq.~\eqref{eq:boost_coefficient_def}. 
We fix $\sqrt{s}/m = 3.3$ and $p/m = 0.7$ and vary $\sigma_k$ over the range $4 \leq \sigma_k/m^2 \leq (\sqrt{s}/m - 1)^2$. Figure~\ref{fig:boost_coeffs} shows $\Bc^\ell_{S0}$ as a function of $\sigma_k$ for several choices of $S$ and $\ell \leq S$. 
For fixed $s$, Eq.~\eqref{eq:cm_momentum} shows that the upper endpoint, $\sigma_k \to (\sqrt{s}-m)^2$, corresponds to the spectator momentum $k \to 0$. 
The key feature of $\Bc^\ell_{S0}$ in this limit is $\Bc^\ell_{S0} \to \delta_{\ell S}$. 
As discussed in Sec.~\ref{sec:boost_coefficients}, when no boost is required to map $\p$ to $\p^\star_k$, no mixing between $\ell \neq S$ components is needed to express $Y_{S\lambda}(\hatpstark)$ in terms of $Y_{\ell\lambda}(\bh{\p})$ in Eq.~\eqref{eq:boost_coefficient_def}. 
More generally, the magnitude of this mixing decreases monotonically as the boost between the two frames becomes weaker.
\begin{figure}
    \centering
    \includegraphics[width=0.45\linewidth]{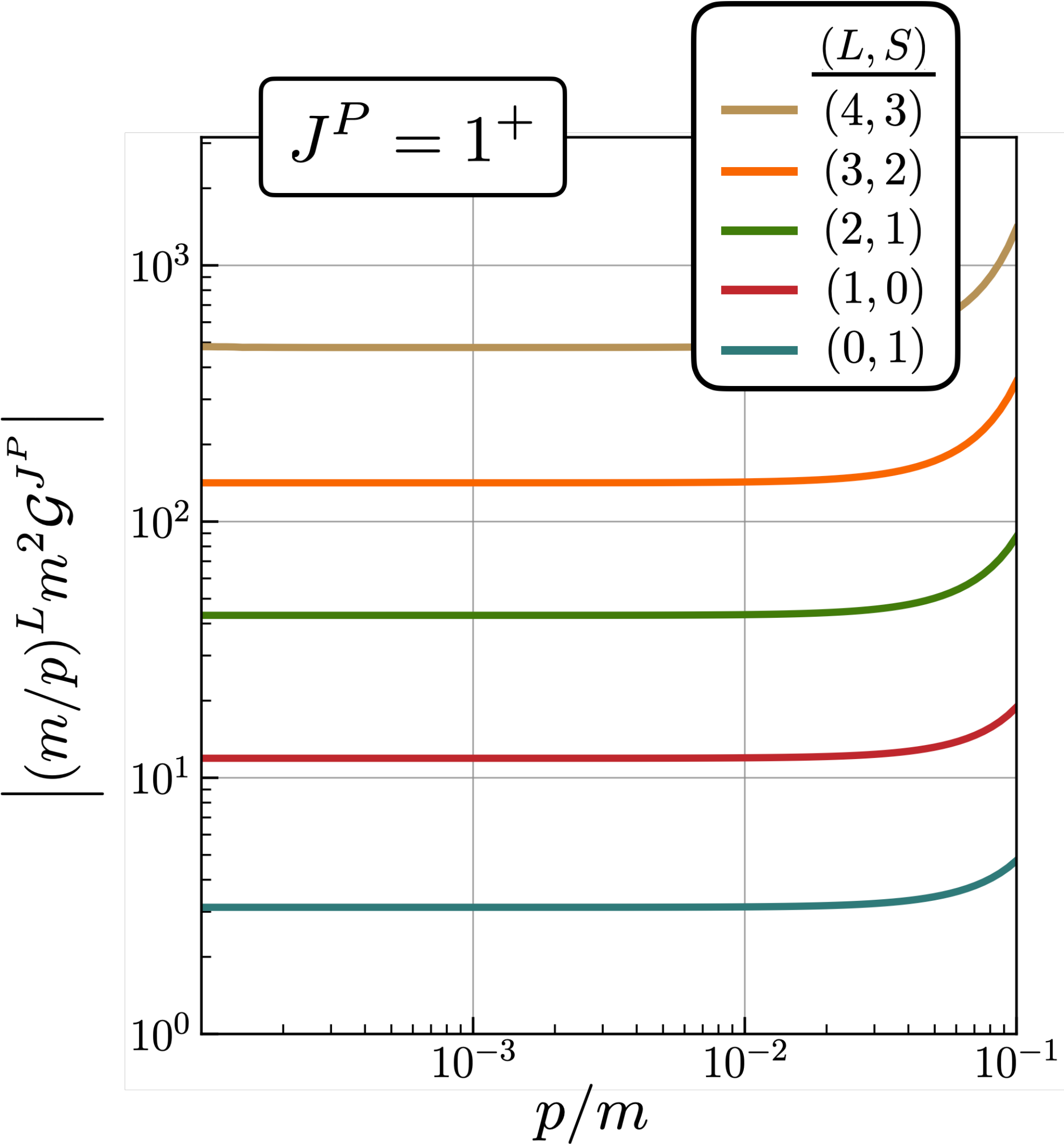}
    \caption{Magnitude of the partial-wave OPE, with expected threshold scale factor $p^{L}$ removed. 
    Pair invariant masses are fixed to $\sqrt{\sigma_p} / m = 2.2$ and $\sqrt{\sigma_k} / m = 2.1$, and $s$ is allowed to vary through $p$, \cf~Eq.~\eqref{eq:cm_momentum}. 
    Incoming and outgoing quantum numbers are set equal, \ie,~$L'=L$ and $S'=S$.}
    \label{fig:ope_threshold_dep}
\end{figure}

Next, we numerically verify the threshold behavior of $\Gc^{J^P}$ derived in Sec.~\ref{sec:threshold}. 
For fixed $\sigma_p$ and $\sigma_k$ above their respective two-body thresholds, the OPE scales as $\Gc^{J^P}_{L'S'\!,\,LS}(p,k) \sim p^{L'}k^L$ as $p,k \to 0$. 
This behavior is exhibited in Fig.~\ref{fig:ope_threshold_dep} for target quantum numbers $J^P = 1^+$, with fixed pair invariant masses $\sqrt{\sigma_k}/m = 2.1$ and $\sqrt{\sigma_p}/m = 2.2$. 
We consider several partial-wave OPE functions with $L=L'$ and $S=S'$, for $L\in\{0,\ldots,4\}$. 
For $L\geq 1$ we choose $S=L-1$, while for $L=0$ we set $S=1$. In each case, we plot $p^{-L'}\Gc^{1^+}$ and observe that it approaches a constant as $p\to0$. 
This numerically supports that the finite-sum expression in Eq.~\eqref{eq:ope_result} has the expected outgoing-threshold scaling, including for higher angular momenta.

\subsection{Functional behavior of the OPE for higher angular momentum}
\label{sec:numerical_functional_behavior}

\begin{figure}
    \centering
    \includegraphics[width=\linewidth]{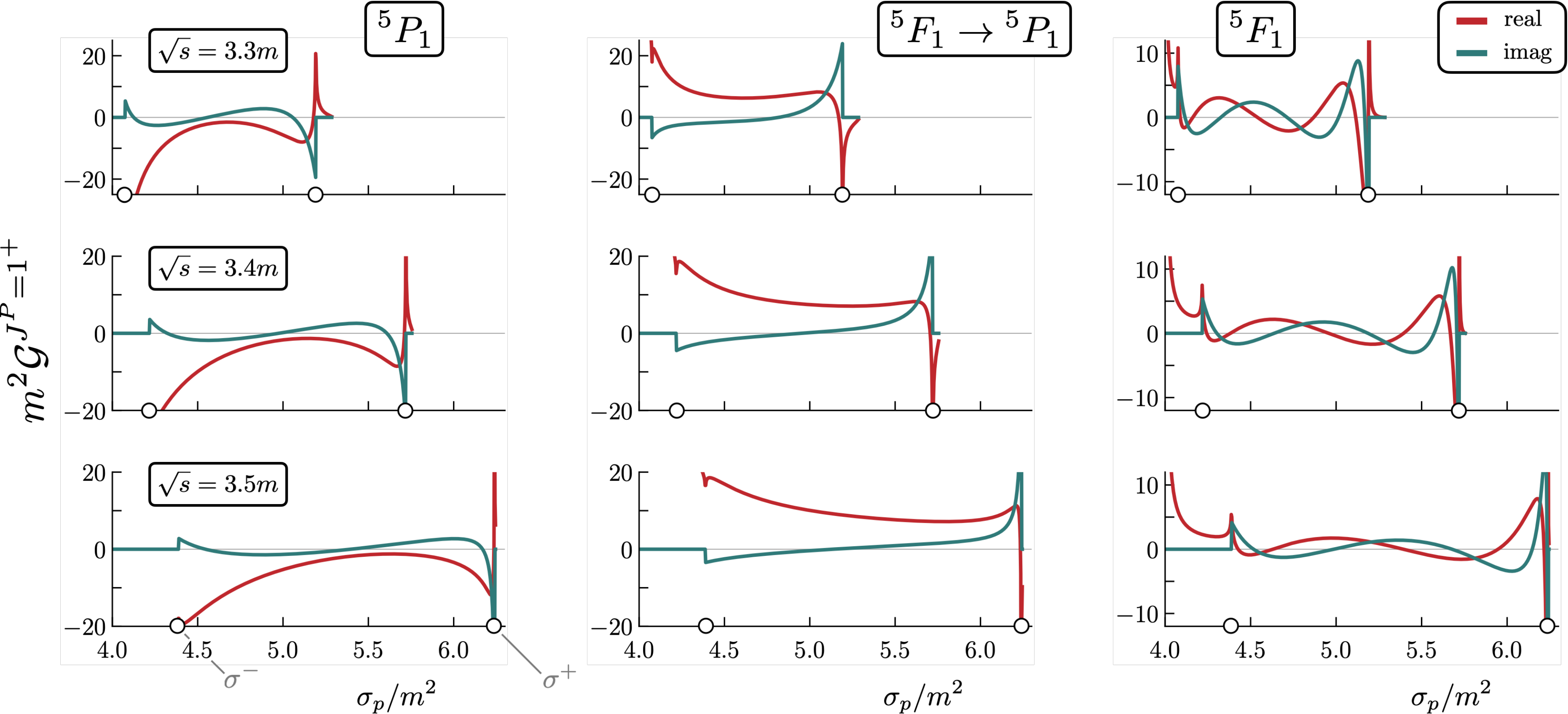}
    \caption{Real and imaginary parts of partial-wave OPE for fixed $J^P = 1^+$ as a function of $\sigma_p / m^2$ for ${}^5P_1$ and ${}^5F_1$ waves. 
    In all plots, $\sqrt{\sigma_k}/m = 2.15$. Each row is plotted with fixed $\sqrt{s}/m = \{3.3, 3.4, 3.5\}$, and each column corresponds to a different set of spin-orbit quantum numbers labeled with spectroscopic notation ${}^{2S+1}L_J$. 
    The branch point locations in $\sigma_p$ are marked.}
    \label{fig:gjp_1p_sigma}
\end{figure}

\begin{figure}
    \centering
    \includegraphics[width=\linewidth]{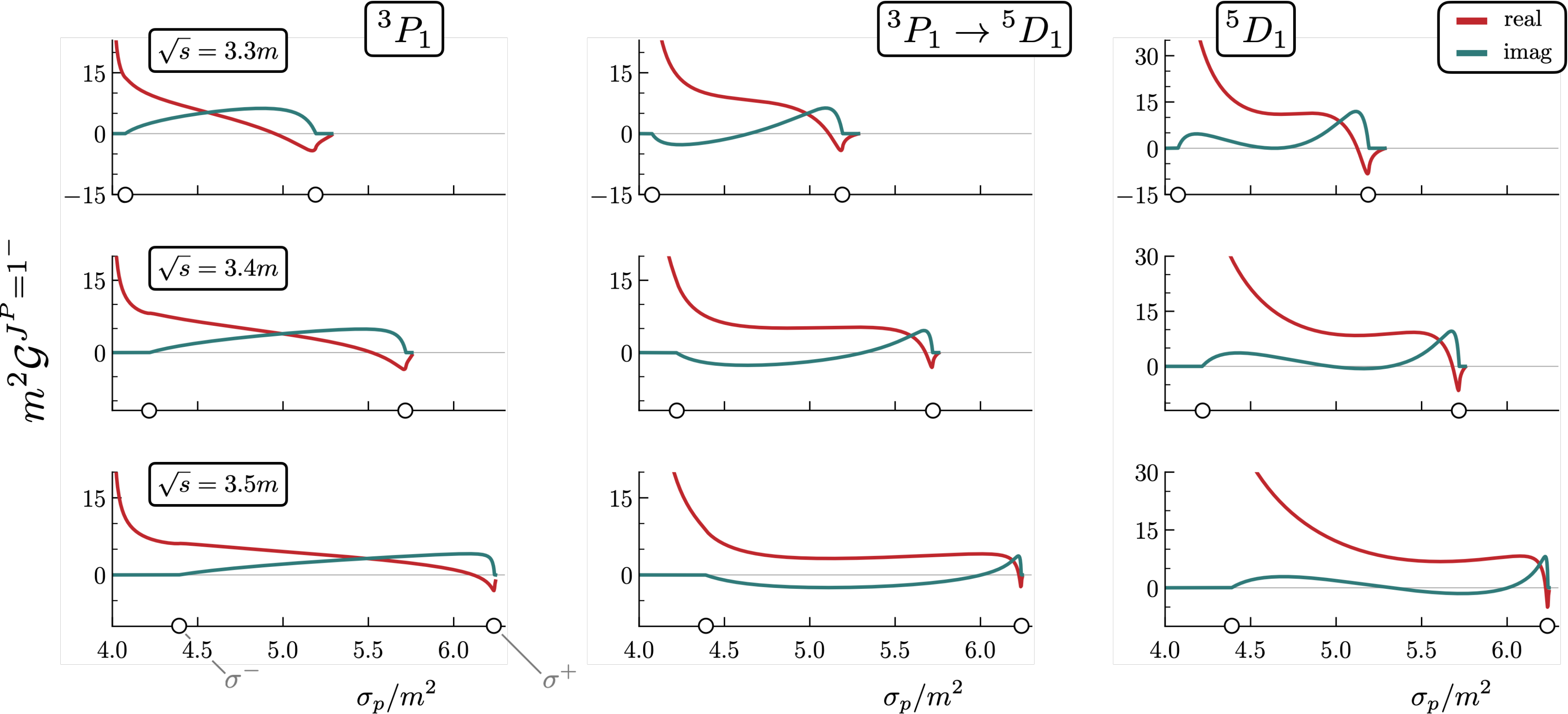}
    \caption{Same as Fig.~\ref{fig:gjp_1p_sigma}, but for $J^P = 1^-$ and ${}^3P_1$ and ${}^5D_1$ waves.}
    \label{fig:gjp_1m_sigma}
\end{figure}

\begin{figure}
    \centering
    \includegraphics[width=\linewidth]{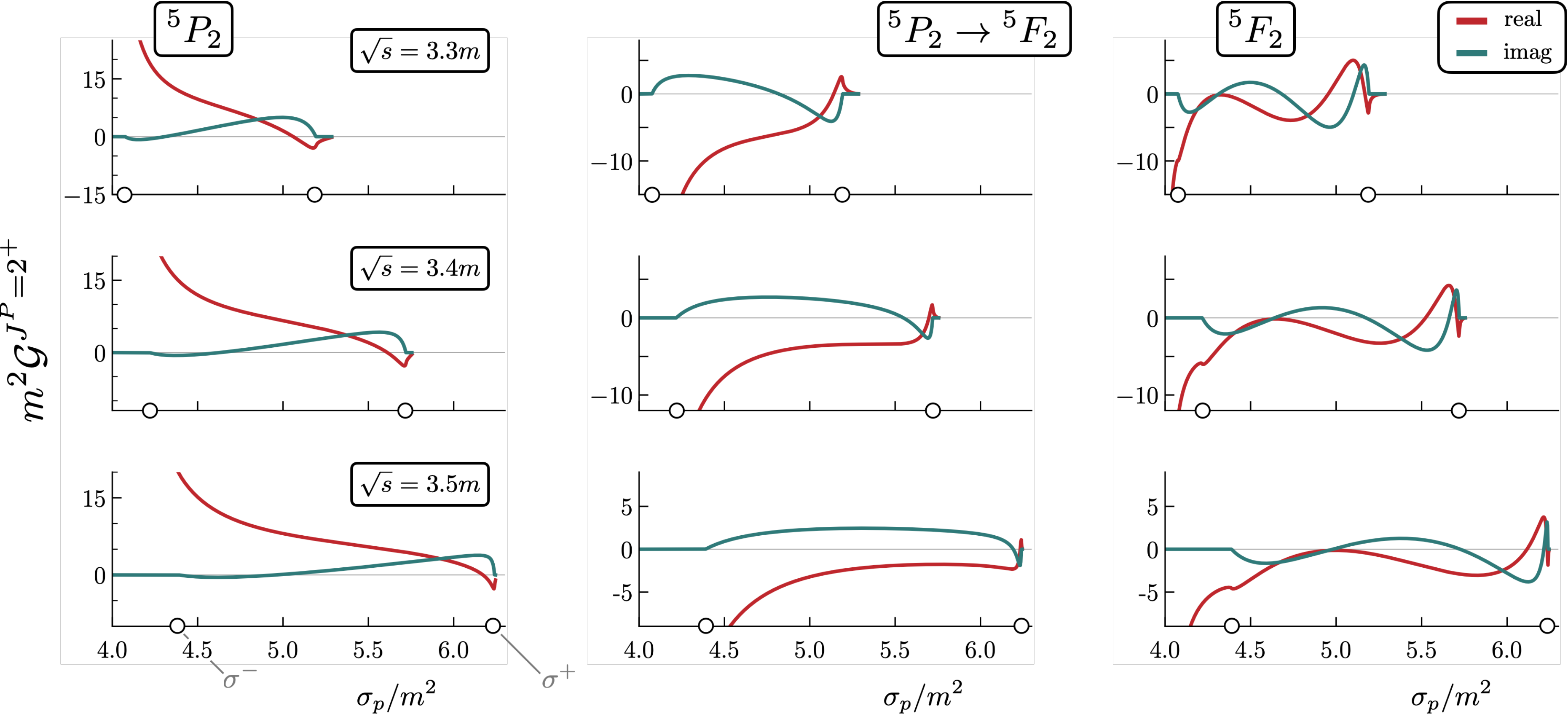}
    \caption{Same as Fig.~\ref{fig:gjp_1p_sigma}, but for $J^P = 2^+$ and ${}^5P_2$ and ${}^5F_2$ waves.}
    \label{fig:gjp_2p_sigma}
\end{figure}

\begin{figure}
    \includegraphics[width=0.45\textwidth]{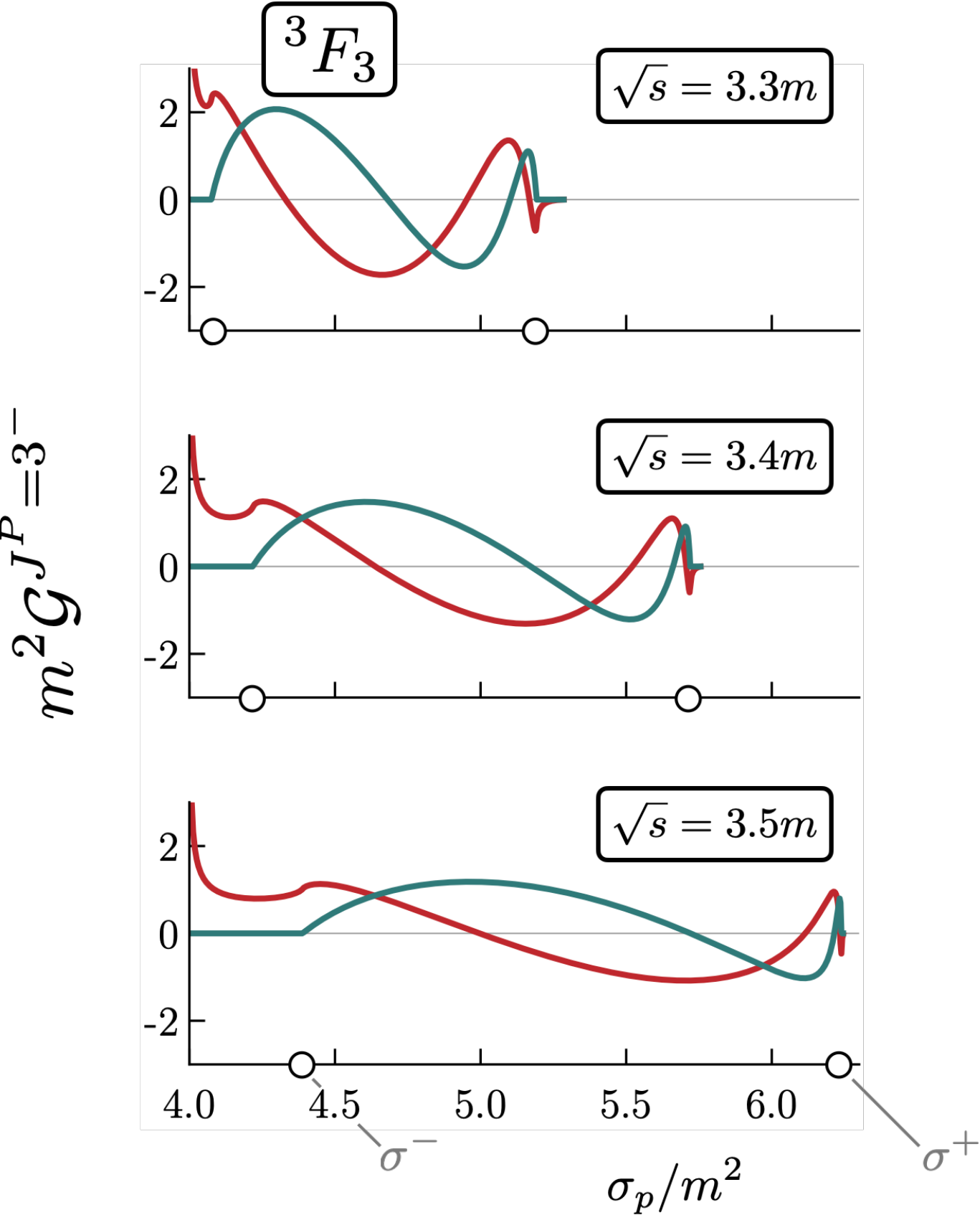}
    \caption{Same as Fig.~\ref{fig:gjp_1p_sigma}, but for $J^P =3^-$ and the ${}^3F_3$ wave.}
    \label{fig:gjp_3m_sigma}
\end{figure}

We illustrate the partial-wave OPE function of Eq.~\eqref{eq:ope_result} for several target quantum numbers $J^P$. Figures~\ref{fig:gjp_1p_sigma}--\ref{fig:gjp_3m_sigma} show its real and imaginary parts at fixed $\sqrt{\sigma_k}/m=2.15$, plotted as functions of $\sigma_p$ over the physical range $4 \leq \sigma_p/m^2 \leq (\sqrt{s}/m-1)^2$. 
The rows correspond to $\sqrt{s}/m = 3.3, 3.4,$ and $3.5$, while the columns show different spin-orbit transitions, ${}^{2S+1}L_J \to {}^{2S'+1}L'_J$, within the chosen $J^P$ sector. 
Diagonal elements are shown with a single spectroscopic label. 
For the interested reader, we present plots of the partial-wave OPE in App.~\ref{app:sec:plots_in_s} that are analogous to Figs.~\ref{fig:gjp_1p_sigma}--\ref{fig:gjp_3m_sigma}, where instead we plot in the $s$ variable and vary $\sigma_p$ across the rows of each figure.

The OPE function inherits branch points from the Legendre functions $Q_j$, as described in Sec.~\ref{sec:summary_results}. 
In our numerical evaluations, we take the physical boundary value by evaluating $Q_j$ just above its branch cut, $\zeta_{pk} \to \zeta_{pk}+i\epsilon$ with $\epsilon>0$. 
Since $\zeta_{pk}$ depends on $s$ and the pair invariant masses, any auxiliary imaginary shifts assigned to these variables must be taken in an ordered limit. 
Following Ref.~\cite{Jackura:2023qtp}, we take $s \to s+i\epsilon_s$ and $\sigma \to \sigma+i\epsilon_\sigma$, with $0<\epsilon_\sigma<\epsilon_s$. 
We then choose the explicit shift of the Legendre-function argument to satisfy $\epsilon>\epsilon_s>\epsilon_\sigma$, so that the boundary value of $Q_j(\zeta_{pk})$ is fixed by the approach from the upper half-plane. 
Marked on each plot in Figs.~\ref{fig:gjp_1p_sigma}--\ref{fig:gjp_3m_sigma} are the locations of the branch points of the OPE function, occurring at the values $\zeta_{pk} = \pm 1$. 
The locations in $\sigma_p$ for fixed $s$ and $\sigma_k$ are given by~\cite{Jackura:2018xnx}
\begin{align}
    \label{eq:ope_boundary}
    \sigma_p^{\pm} = \frac{1}{2}\left(3m^2 + s - \sigma_k \pm \lambda^{1/2}(s, m^2, \sigma_k) \sqrt{1 - \frac{4m^2}{\sigma_k}}\right) \, .
\end{align}
The broad analytic features across the OPE plots in Figs.~\ref{fig:gjp_1p_sigma}--\ref{fig:gjp_3m_sigma} are governed by S matrix unitarity, which constrains the imaginary part of the amplitude to be nonzero in the region of physical one-particle exchange, $\zeta_{pk} \in [-1, 1]$ or $\sigma_p \in [\sigma_p^{-},\sigma_p^{+}]$. 

In a three-pion system, the spin-orbit basis contains infinitely many allowed angular-momentum channels once the appropriate isospin and Bose-symmetry assignments are included. 
In Ref.~\cite{Jackura:2023qtp}, examples were shown up to $S=1$ and $L=2$, corresponding to $P$ waves in the pair subsystem and $D$ waves in the pair-spectator system. 
Here we demonstrate that the finite-sum representation, Eq.~\eqref{eq:ope_result}, can be applied straightforwardly to higher angular momenta, showing representative OPE amplitudes with pair angular momentum up to $S=2$ and pair-spectator orbital angular momentum up to $L=3$ in Figs.~\ref{fig:gjp_1p_sigma}--\ref{fig:gjp_3m_sigma}. 

The examples in Figs.~\ref{fig:gjp_1p_sigma} and~\ref{fig:gjp_1m_sigma} include channels relevant to $3\pi$ amplitude analyses in $J^P = 1^+$ and $1^-$, respectively. 
For the $J^P=1^+$ system, there is evidence that the inclusion of $\pi\pi$ $D$ waves such as those shown in Fig.~\ref{fig:gjp_1p_sigma} would be sensible, as it was reported in Ref.~\cite{CLEO:1999rzk} that a fit of $\tau \to \nu_\tau \pi^- \pi^0 \pi^0$ data demonstrated a statistically significant signal for the $a_1(1260)$ resonance to decay into a ${}^5 P_1$ final state --- in which the $\pi \pi$ subsystem resonates into the $f_2(1270)$. 
Figure~\ref{fig:gjp_1m_sigma} shows the target quantum numbers $J^P = 1^-$, relevant for systems such as the spin-exotic $\pi_1(1600)$ resonance~\cite{COMPASS:2015gxz,COMPASS:2018uzl,COMPASS:2020yhb}, with the inclusion of ${}^3P_1$ and ${}^5D_1$ waves. 
Figures~\ref{fig:gjp_2p_sigma} and~\ref{fig:gjp_3m_sigma} similarly illustrate the higher-spin sectors needed for spin-2 and spin-3 resonance studies, such as the $a_2(1320)$ and $\omega_3(1670)$, respectively. 
Our purpose here is not to perform the full isospin-coupled phenomenological analysis, but to show that the OPE kernel itself is readily evaluated for the required higher-angular-momentum spin-orbit channels.

All plots clearly demonstrate consistency with unitarity as applied to the OPE, with $\im \Gc^{J^P} \neq 0$ in the region $\sigma_p \in [\sigma^-_p, \sigma^+_p]$. 
In each row of the figure, we increase $\sqrt{s}$ and observe that the physical OPE region grows according to Eq.~\eqref{eq:ope_boundary}.
Note the diverging behavior of the OPE, $\Gc^{J^P}_{L'S'LS} \to (q^\star_p)^{-S'}$ as $\sigma_p \to 4m^2$. 
As shown in Sec.~\ref{sec:threshold}, this divergence will not produce divergent behavior in the full amplitude $\Mc^{J^P}_3$. 
Within the integral equation construction of $\Mc^{J^P}_3$, $\Gc^{J^P}$ is always accompanied with two factors of the two-body partial-wave amplitude (one each for initial/final pairs) of spins $S$ and $S'$, whose exchanged momenta scale respectively as $\Mc_2 \to (q^\star)^{S}$ and $(q^\star)^{S'}$ in the same limit and cancel the threshold divergences within the OPE.

It is worth comparing the features of the OPE in Fig.~\ref{fig:gjp_1m_sigma} vs. Fig.~\ref{fig:gjp_1p_sigma}.
For $J^P=1^-$, the number of critical points of the OPE function within the physical OPE region is less than what is naively expected as compared with $J^P = 1^+$. 
In both Figs.~\ref{fig:gjp_1p_sigma} and \ref{fig:gjp_1m_sigma}, $\jmaxcg = J + S + S' = 1 + 2 + 2 = 5$, so we would expect the polynomial behavior for $J^P = 1^-$ to reflect that of $J^P = 1^+$. 
Instead, these figures indicate that differing parities at a given total $J$ may have differing polynomial order in $\zeta_{pk}$, and we show in App.~\ref{app:sec:parity_suppression} that the largest value of $j$ that contributes to the sum in Eq.~\eqref{eq:ope_result} is $\jmaxcg-1$ for $J^P$ with natural parity choices $P = (-1)^{L+S + 1} = (-1)^J$ (for systems of three pseudoscalars).
Finally, note that the OPE singularities are suppressed in this channel. As discussed in Sec.~\ref{sec:derivation}, the $Q_j$ functions can be written in terms of $Q_0$, which results in a partial-wave OPE function as shown in Eq.~\eqref{eq:ope_split_form}. 
For some $J^P$ choices, we find that $\Tc^{J^P} \to 0$ polynomially as $\zeta_{pk} \to \pm 1$, and therefore the logarithmic singularity of the OPE is softened in these cases, as observed in Ref.~\cite{Jackura:2023qtp}. 
%

\subsection{Relative contribution of subchannel partial waves}
\label{sec:relative_contribution}

The waves presented here represent only a small subset of the infinite number of partial waves available for a given $J^P$. 
In practice, any reconstruction of the partial-wave three-body amplitude $\Mc_3^{J^P}$ from data must truncate this space of partial waves, whether that data be from lattice QCD through the L\"uscher method or experimental spin-density matrix elements. 
As a final numerical test, we construct the full OPE matrix over a range of physical and unphysical kinematics using wavesets that contain all allowed partial-wave channels through a specified angular-momentum truncation. 
We then compare the relative contributions of the individual channels as the truncation is varied.
\begin{figure}
    \centering
    \includegraphics[width=0.65\linewidth]{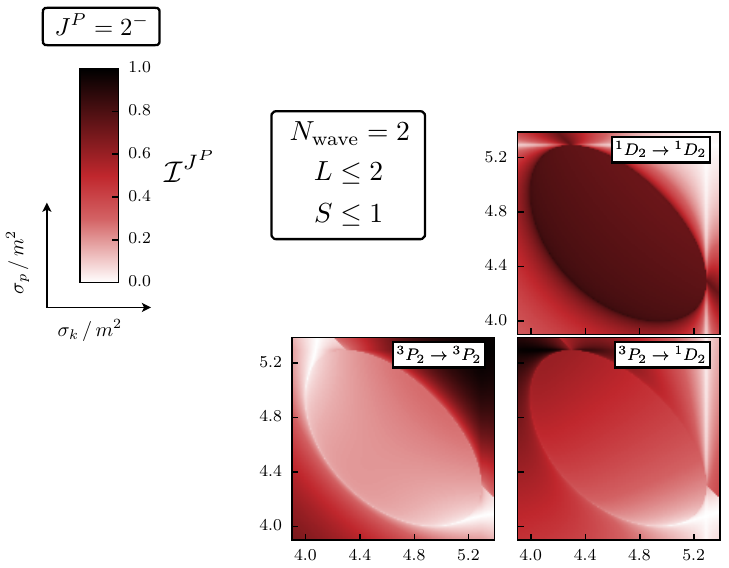}
    \caption{Normalized magnitude of the OPE in the $J^P=2^-$ sector for a waveset with $L\le 2$ and $S\le 1$, yielding $N_{\mathrm{wave}} = 2$ waves. 
    The intensity is shown as a function of $\sigma_k$ and $\sigma_p$.}
    \label{fig:ope_norm_2m_trunc1}
\end{figure}

\begin{figure}
    \centering
    \includegraphics[width=0.9\linewidth]{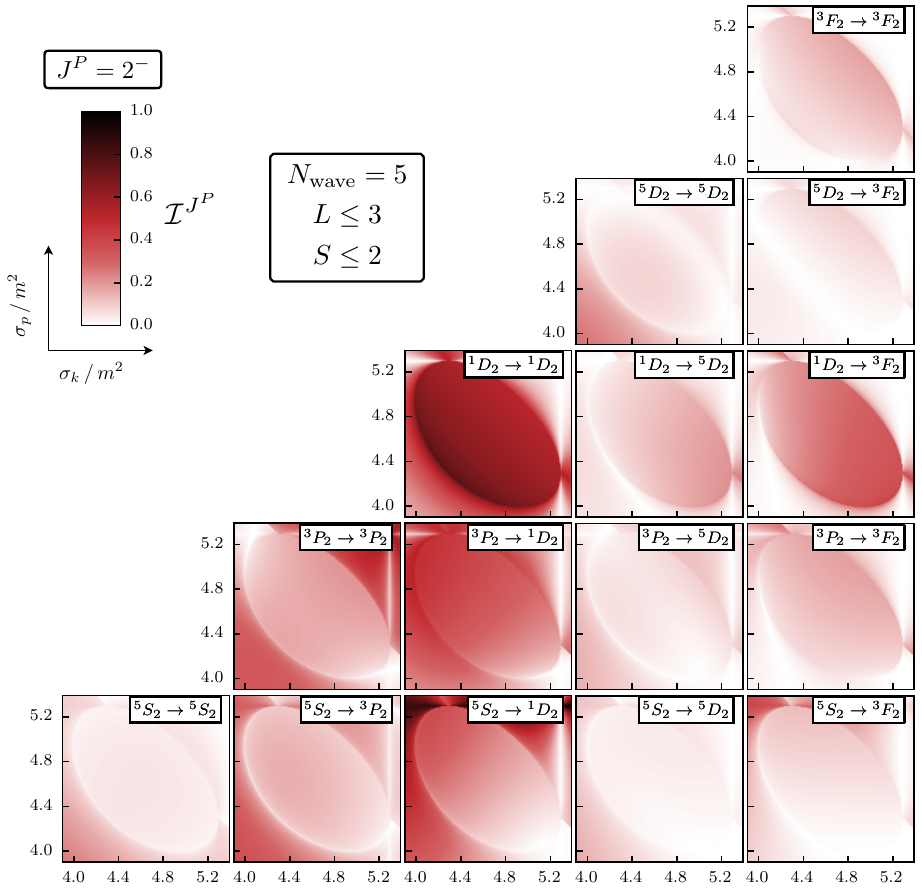}
    \caption{Same as Fig.~\ref{fig:ope_norm_2m_trunc1} but for a trunctation $L\le 3$ and $S\le 2$, with $N_{\mathrm{wave}} = 5$ total waves.}
    \label{fig:ope_norm_2m_trunc2}
\end{figure}

\begin{figure}
    \centering
    \includegraphics[width=0.99\linewidth]{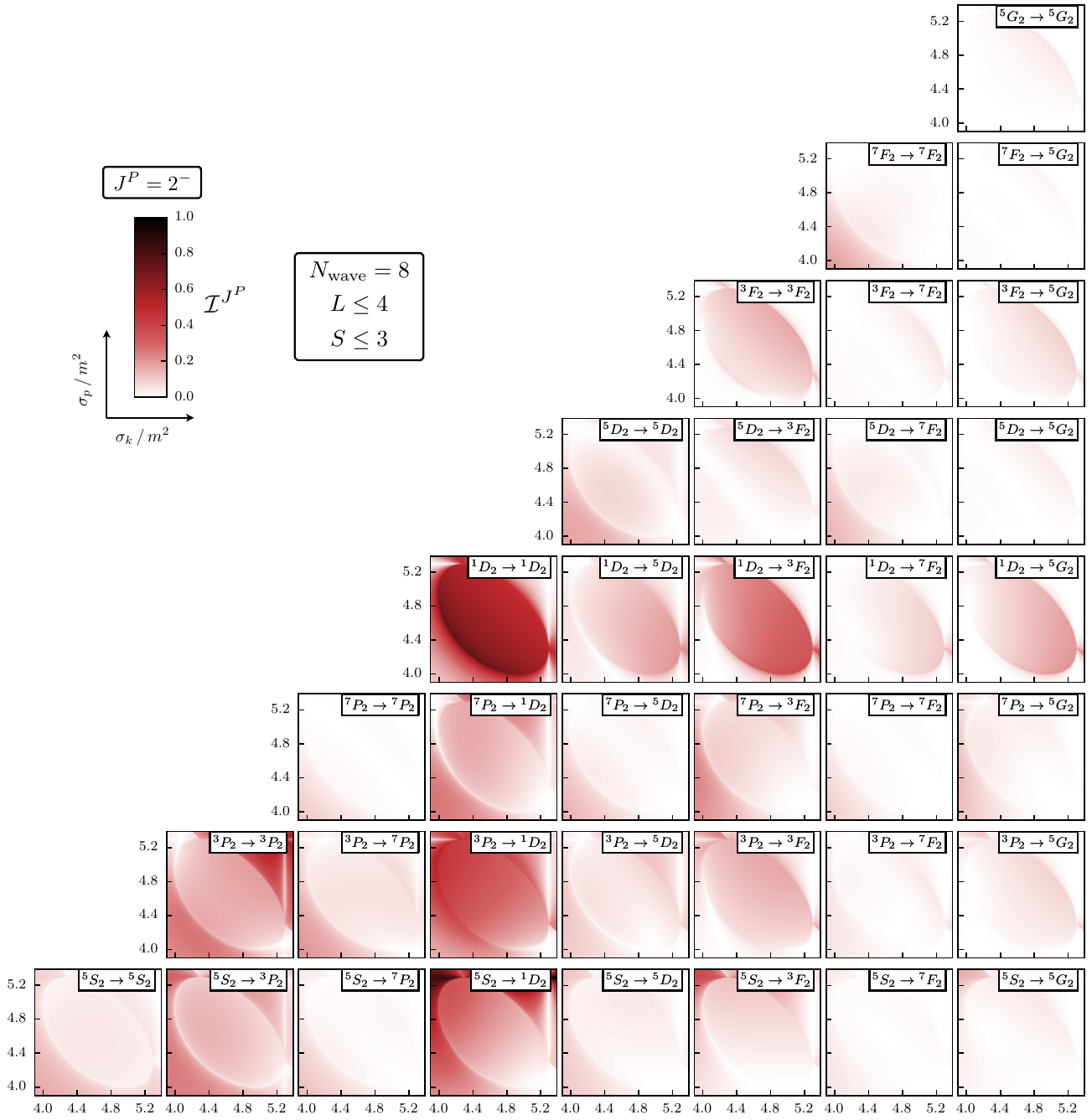}
    \caption{Same as Fig.~\ref{fig:ope_norm_2m_trunc1} but for a trunctation $L\le 4$ and $S\le 3$, with $N_{\mathrm{wave}} = 8$ total waves.}
    \label{fig:ope_norm_2m_trunc3}
\end{figure}

To illustrate the impact of truncation, we examine the OPE as a matrix in spin-orbit subchannel space. 
To properly compare partial waves with different threshold scaling, we examine a modified OPE matrix element,
\begin{align}
    \tilde{\Gc}^{J^P}_{L'S',LS}(p,k) = \left(\frac{\qstarp}{m}\right)^{S'} \Gc^{J^P}_{L'S',LS}(p,k) \left(\frac{\qstark}{m}\right)^{S} 
\end{align}
We then define the intensity $\Ic^{J^P}$ as the normalized magnitude of the modified OPE,
\begin{align}
    \Ic_{L'S',\,LS}^{J^P}(p,k) 
    = 
    \left(
    \sum_{\bar{L}',\bar{L} = 0}^{L_{\max}} 
    \sum_{\bar{S}',\bar{S} = 0}^{S_{\max}}
    \Big\lvert \, \tilde{\Gc}_{\bar{L}'\bar{S}',\,\bar{L}\bar{S}}^{J^P}(p,k)\Big\rvert^2 
    \right)^{-1/2} \, 
    \Big\lvert \, 
    \tilde{\Gc}_{L'S',\,LS}^{J^P}(p,k) 
    \, \Big\rvert \, ,
    \label{eq:normalized_ope}
\end{align}
where $L_{\max}$ and $S_{\max}$ denote the maximum values retained in the truncation scheme. 
Note the sums are implicitly restricted by angular momentum selection rules to those values which contribute to the target $J^P$ below the truncation. 
For a waveset of size $N_{\mathrm{wave}}$,~\footnote{Explicitly, $N_{\textrm{wave}}$ is the number of unique $(L,S)$ combinations contributing to total $J^P$ at the chosen truncation level.} the magnitude is normalized by summing over all $N_{\mathrm{wave}} \times N_{\mathrm{wave}}$ matrix elements for a given $J^P$. 
In Figs.~\ref{fig:ope_norm_2m_trunc1} through~\ref{fig:ope_norm_2m_trunc3}, we show the application of Eq.~\eqref{eq:normalized_ope} for $J^P = 2^-$ at three different truncation levels. 
Since the effective matrix is symmetric about the anti-diagonal in the figure layout, only the lower-right triangle is shown.

In Fig.~\ref{fig:ope_norm_2m_trunc1}, we show $\Ic^{J^P = 2^-}$ at fixed $\sqrt{s}/m = 3.3$ for $L\le 2$ and $S \le 1$, corresponding to $N_{\mathrm{wave}} = 2$ and the two waves ${}^3P_2$ and ${}^1D_2$. 
The physical region in both $\sigma_k$ and $\sigma_p$ is again $4m^2 \le \sigma_k \le (\sqrt{s}-m)^2$, and the upper bound is therefore $(\sqrt{s}/m-1)^2 = 5.29$. 
The color gradient represents $\mathcal I^{J^P}$ and shows that, within this truncation, the ${}^1D_2\to{}^1D_2$ contribution is dominant in the physical region. 
A boundary corresponding to the Kibble cubic is clearly visible in each panel; this boundary separates the physical region of the OPE, defined by $\Phi(p,k) \ge 0$, with
\begin{align}
    \Phi(p,k) 
    = 
    \sigma_p\sigma_k(s+3m^2 - \sigma_p-\sigma_k) 
    - 
    m^2(s - m^2)^2,
\end{align}
as discussed in Ref.~\cite{Jackura:2023qtp} and references therein. 
Outside this region, the OPE may contain additional unphysical singularities that arise from our prescription for defining the OPE. In particular, the lines of vanishing intensity at large $\sigma_k$ and $\sigma_p$ correspond to the zero-momentum points $k = 0$ and $p = 0$, respectively.

Relaxing the truncation level to $L\le 3$ and $S\le 2$ now gives $N_{\mathrm{wave}} = 5$ waves, shown in Fig.~\ref{fig:ope_norm_2m_trunc2}. 
Note that the intensity scale changes relative to Fig.~\ref{fig:ope_norm_2m_trunc1}, since the normalization depends on the number of waves included. 
Again, the ${}^1D_2 \to{}^1D_2 $ element dominates the set, and we generally find that the diagonal entries become less intense for higher $S$. 
Surprisingly, the off-diagonal entries that overlap with a ${}^1D_2$ wave contribute substantially to the waveset compared to lower-order subchannels --- compare for instance ${}^1D_2 \to {}^3F_2$ to ${}^3P_2 \to {}^3P_2$. 
This example highlights the need for carefully considering the effects of truncating the spin-orbit subchannel space. 
In a practical three-body partial-wave analysis, a truncation at the level $L_{\textrm{max}} = S_{\textrm{max}} = 2$ would neglect this ${}^1D_2 \to {}^3F_2$ component, which could introduce a significant systematic error in the $J^P=2^-$ amplitude.~\footnote{Note that we only consider the OPE contribution to $\Ic^{J^P}$, whereas in a physical intensity distribution, the OPE would be accompanied by two factors of the two-body subchannel amplitude $\Mc_2$ as described in Sec.~\ref{sec:threshold}. 
The interaction strength of the two-body subchannels would impact the partial-wave truncation error independently of the kinematic arguments we make in this section.} 
This unintuitive strength of a higher-order partial wave is explained by the fact that Figs.~\ref{fig:ope_norm_2m_trunc1}--\ref{fig:ope_norm_2m_trunc3} do not depict a threshold limit in one single kinematic variable with others fixed, like in Sec.~\ref{sec:threshold}. 
Instead, we present the partial-wave OPE as a distribution in multiple variables, where threshold behavior in differing kinematic limits produces non-uniform scaling across the spin-orbit waveset.
Unlike in two-body scattering, where partial waves $\Mc_2^S(\sigma_k)$ scale near threshold as $(\qstark)^{2S}$ at low energy, there is no clear power-counting scheme that predicts which three-body subchannel waves have the greatest strength when multiple kinematic variables are free to vary.

Finally, to showcase the versatility of the finite-sum OPE in handling a large waveset, we include waves with $L\le 4$ and $S \le 3$ in Fig.~\ref{fig:ope_norm_2m_trunc3}, giving $N_{\textrm{wave}} = 8$. 
Consistent with Fig.~\ref{fig:ope_norm_2m_trunc2}, entries overlapping with ${}^1D_2$ dominate the set, further supporting the need for a careful analysis of truncation in $L$ and $S$.
For example, neglecting $F$-wave contributions to the pair subsystems excludes the ${}^7P_2\to{}^1D_2$ entry, which is of greater magnitude than ${}^5S_2 \to {}^5S_2$. 
Again, no straightforward $(L,S)$-power-counting scheme predicts \emph{a priori} that, for instance, ${}^7P_2 \to {}^1D_2$ is comparable to ${}^3F_2 \to {}^3F_2$ and ${}^5S_2 \to {}^3P_2$ for these kinematics. 
The influence of the OPE across a broad range of spin-orbit subchannels was also identified in Ref.~\cite{PitangaLachini:2026lyd}, where the authors perform a lattice QCD study of the $T_{cc}$ in $DD^\star$ scattering. 
In that work, the authors found that by including the OPE in their amplitude parameterization of the $T_{cc}$ channel, they were able to reduce their $\Kc$-matrix parameter set across multiple partial waves and still produce a significantly better fit to their finite-volume $DD^*$ spectra. 
Phenomenological studies of three-meson resonance production also find that changing the size of the waveset will significantly impact the quality of fits to three-body mass spectra~\cite{COMPASS:2015gxz,COMPASS:2018uzl,COMPASS:2020yhb}. 
The results of this section further support the growing consensus within the hadron spectroscopy community that an extensive spin-orbit waveset is necessary to fully assess systematics in determining three-body partial-wave amplitudes.

\section{Summary}
\label{sec:summary}

We have derived a finite-sum expression for the partial-wave OPE, a single formula that is valid for arbitrary total angular momentum $J$ and arbitrary spin-orbit subchannels. 
A key step in the derivation is to express the boosted angles of the spin-helicity matrix $\Hc$ in terms of CM frame kinematics, facilitated by a set of boost coefficients $\Bc$. 
The rank-one boost coefficients are then defined to reproduce the Lorentz transformation of $\p$ to $\p^\star_k$, given by Eqs.~\eqref{eq:recap_boost_par} and \eqref{eq:recap_boost_perp}. 
Using the rank-one coefficients as a base case, we give a general formula for $\Bc$ in terms of lower-rank coefficients in Eqs.~\eqref{eq:boost_coeff_recursive} and \eqref{eq:boost_coeff_product}. 
Finally, the familiar recoupling and orthogonality properties of the spherical harmonics allow us to simplify the partial-wave projection of $\Gc$. 
The main result, summarized in Eqs.~\eqref{eq:ope_result}--\eqref{eq:boost_coeff_recursive}, is ready for application to lattice QCD and phenomenological three-body amplitude analyses and easily manages the complexity of partial-wave subchannels at higher total angular momentum. 
This universal analytic representation for the partial-wave OPE also allowed us to prove in Sec.~\ref{sec:threshold} the anticipated threshold scaling in various kinematic limits.

We carried out numerical checks verifying this threshold behavior as well as the expected singular behavior at the boundaries of the physical OPE region, both features consistent with unitarity. 
In Sec.~\ref{sec:relative_contribution}, we show that the spin-orbit basis does not establish a straightforward hierarchy of important partial waves across multiple variables at low energies. 
This supports the need for a robust waveset in three-body spectroscopy studies, in agreement with the findings in phenomenological studies. 
While we have only shown a small selection of $J^P$ systems, we have also computed and verified results up through $J^P = 6^{\pm}$ with wave content up through $G$ waves in both $L$ and $S$. 
We conclude for reasonable $J^P$ systems relevant for hadron spectroscopy, our finite-sum representation for the partial-wave OPE produces stable and reliable numerical results. 

Building on the analytic framework of Refs.~\cite{Jackura:2023qtp,Briceno:2024ehy}, the formula presented here provides a more compact and scalable representation of the partial-wave OPE for three spinless particles. By eliminating case-by-case coefficient calculations, it facilitates both enlarged wavesets at fixed $J^P$ and analyses of spin-two and higher-spin resonances. Although the present derivation assumes spinless particles, the underlying projection strategy may prove useful in extending the $\3\to\3$ formalism to particles with intrinsic spin.

\begin{acknowledgments}

We thank Sebastian Dawid for useful discussions. 
N.C.C. acknowledges support from the Virginia Space Grant Consortium Graduate Research STEM Fellowship Program, and from the Jefferson Science Associates (JSA) Graduate Fellowship through the JSA/JLab Graduate Fellowship Program. 
N.C.C. also acknowledges support by the U.S. Department of Energy (DOE), Office of Science, Office of Workforce Development for Teachers and Scientists, Office of Science Graduate Student Research (SCGSR) program, during the period in which part of this work was carried out. 
The SCGSR program is administered by the Oak Ridge Institute for Science and Education for the DOE under contract number DE‐SC0014664. 
A.W.J. acknowledges support by the National Science Foundation (NSF) under Grant No. PHY-2609972. 
A.W.J. also acknowledges the support of the U.S. DOE ExoHad Topical Collaboration, contract DE-SC0023598.

\end{acknowledgments}

\appendix

\section{Quantum angular momentum identities}
\label{app:sec:ang_mom}

In this appendix, we collect the angular momentum identities used throughout this work. We follow the conventions of Ref.~\cite{VMK} for Clebsch-Gordan coefficients, Wigner $D$ matrices, and spherical harmonics.

The Clebsch-Gordan coefficients define the unitary transformation between coupled and uncoupled angular-momentum bases,
\begin{align}
	\Cc^{jm_j}_{j_1m_1,\,j_2m_2}
	&\equiv
	\braket{jm_j|j_1m_1,\,j_2m_2} \, ,
	\label{eq:app_cg_convention}
\end{align}
with the Condon-Shortley phase convention~\cite{VMK}. 
The state $\ket{jm}$ represents the usual SU(2) angular momentum eigenstate with total angular momentum $j$ and $z$-projection $m$ with $|m|\le j $, and $\ket{j_1m_1,\,j_2m_2}$ is the usual tensor product of two such states. 
All angular momenta appearing in this work are integers. 
Consequently, factors such as $(-1)^{2k}$ are unity for any angular-momentum quantum number $k$, which simplifies several phase relations below.
Equation~\eqref{eq:app_cg_convention} vanishes unless the selection rules, $m = m_1+m_2$ and $\lvert j_1-j_2\rvert \le j \le j_1+j_2$, are satisfied.

The Clebsch-Gordan coefficients respect numerous symmetry relations. 
For this work in particular, we rely on
\begin{subequations}
\begin{align}
	\Cc^{jm}_{j_1m_1,\,j_2m_2}
	&=
	(-1)^{j_1+j_2-j}\,
	\Cc^{j,-m}_{j_1,-m_1,\,j_2,-m_2} \, ,
	\label{eq:app_cg_sym_1}
	\\[5pt]
	&=
	(-1)^{j_1+j_2-j}\,
	\Cc^{jm}_{j_2m_2,\,j_1m_1} \, ,
	\label{eq:app_cg_sym_2}
	\\[5pt]
	&=
	(-1)^{j_2+m_2} \,
	\sqrt{\frac{2j+1}{2j_1+1}}\,
	\Cc^{j_1,-m_1}_{j,-m,\,j_2m_2} \, .
	\label{eq:app_cg_sym_3}
\end{align}
\end{subequations}
Combining Eqs.~\eqref{eq:app_cg_sym_1} with~\eqref{eq:app_cg_sym_3} for the special case of integer quantum numbers and $m_2 = 0$ gives a useful permutation identity
\begin{align}
	\Cc^{jm}_{j_1m,\,j_20}
	&=
	(-1)^{j_1-j}
	\sqrt{\frac{2j+1}{2j_1+1}}\,
	\Cc^{j_1m}_{jm,\,j_20} \, ,
	\label{eq:app_cg_zero_projection_permutation}
\end{align}
where the factor $(-1)^{2j_2} = 1$ since $j_2$ is an integer. 
The orthogonality relations are
\begin{subequations}
\begin{align}
	\sum_{m_1,m_2}
	\Cc^{jm}_{j_1m_1,\,j_2m_2} \,
	\Cc^{j'm'}_{j_1m_1,\,j_2m_2}
	&=
	\delta_{jj'}\delta_{mm'} \, ,
	\label{eq:app_cg_orthogonality_1}
	\\[5pt]
	\sum_{j,m}
	\Cc^{jm}_{j_1m_1,\,j_2m_2} \,
	\Cc^{jm}_{j_1m'_1,\,j_2m'_2}
	&=
	\delta_{m_1m'_1}\delta_{m_2m'_2} \, .
	\label{eq:app_cg_orthogonality_2}
\end{align}
\end{subequations}
We also use the partial orthogonality relations,
\begin{subequations}
\begin{align}
	\label{eq:partial_orthogonality_1}
	\sum_{m,m_1} 
	\Cc^{j m}_{j_1 m_1, \, j_2 m_2} \,
	\Cc^{jm}_{j_1m_1,\, j_2 m_2'}
	& = 
	\frac{2j+1}{2j_2+1} \,
	\delta_{m_2m_2'} \, ,
	\\[5pt]
	\label{eq:partial_orthogonality_2}
	\sum_{m,m_2} 
	\Cc^{j m}_{j_1 m_1, \, j_2 m_2} \,
	\Cc^{jm}_{j_1m_1',\, j_2 m_2}
	& = 
	\frac{2j+1}{2j_1+1} \,
	\delta_{m_1 m_1'} \, .
\end{align}
\end{subequations}

Wigner $D$ matrices are matrix elements of active rotations in the $\ket{jm}$ basis. 
We denote them by $D^{(j)}_{m'm}(\bh{\n})$, where $\bh{\n}$ has polar and azimuthal angles $(\theta_n,\varphi_n)$. 
Throughout this work, our notation abbreviates the Euler-angle convention
$D^{(j)}_{m'm}(\bh{\n})\equiv D^{(j)}_{m'm}(\varphi_n,\theta_n,0)$~\cite{VMK}.

For a common, fixed second index $n$, the $D$ matrices respect the orthogonality relation
\begin{align}
	\int\!\diff\hat{\n}\,
	D^{(j)\,*}_{mn}(\bh{\n})\,
	D^{(j')}_{m'n}(\bh{\n})
	&=
	\frac{4\pi}{2j+1}
	\delta_{jj'}\delta_{mm'} \, ,
	\label{eq:app_cg_D_orthogonality}
\end{align}
where the measure is $\diff\bh{\n} \equiv \diff\varphi_n\,\diff\cos\theta_n$, and the integral is taken over the entire solid angle, $0\le \varphi_n\le 2\pi$ and $-1\le \cos\theta_n \le +1$. 
Here $*$ denotes complex conjugation. 
The product of two Wigner $D$ matrices is reduced by the Clebsch-Gordan series
\begin{align}
	D^{(j_1)}_{m_1n_1}(\bh{\n})\,
	D^{(j_2)}_{m_2n_2}(\bh{\n})
	&=
	\sum_{j,m,n}
	\Cc^{j\,m}_{j_1m_1,\,j_2m_2} \,
	\Cc^{j\, n}_{j_1n_1,\,j_2n_2} \,
	D^{(j)}_{mn}(\bh{\n}) \, .
	\label{eq:app_cg_D_product}
\end{align}
Likewise, the inverse of the Clebsch-Gordan series is found by using the orthogonality relation Eq.~\eqref{eq:app_cg_orthogonality_1} on~\eqref{eq:app_cg_D_product},
\begin{align}
	D^{(j)}_{mn}(\bh{\n}) 
	= 
	\sum_{m_1,m_2}\sum_{n_1,n_2} 
	\Cc^{jm}_{j_1m_1,\,j_2m_2}\,
	\Cc^{jn}_{j_1n_1,\,j_2n_2}\,
	D^{(j_1)}_{m_1n_1}(\bh{\n})\,
	D^{(j_2)}_{m_2n_2}(\bh{\n})\,.
\end{align}
By combining both Eqs.~\eqref{eq:app_cg_D_orthogonality} and~\eqref{eq:app_cg_D_product}, we can evaluate the integral over the product of three $D$ matrices,
\begin{align}
	\int\!\diff\bh{\n}\,
	D^{(j)\,*}_{mn}(\bh{\n})\,
	D^{(j_1)}_{m_1n_1}(\bh{\n})\,
	D^{(j_2)}_{m_2n_2}(\bh{\n})
	&=
	\frac{4\pi}{2j+1}
	\Cc^{jm}_{j_1m_1,\,j_2m_2}
	\Cc^{jn}_{j_1n_1,\,j_2n_2} \, .
	\label{eq:app_cg_three_D_integral}
\end{align}
Similarly, an integral over four $D$ matrices follows by first coupling two of the unconjugated $D$ matrices using Eq.~\eqref{eq:app_cg_D_product}, and then using the three-$D$ result above, Eq.~\eqref{eq:app_cg_three_D_integral}. 
Explicitly,
\begin{align}
	&\int\!\diff\bh{\n}\,
	D^{(j)\,*}_{mn}(\bh{\n})\,
	D^{(j_1)}_{m_1n_1}(\bh{\n})\,
	D^{(j_2)}_{m_2n_2}(\bh{\n})\,
	D^{(j_3)}_{m_3n_3}(\bh{\n})
	\nn\\[5pt]
	&\qquad =
	\frac{4\pi}{2j+1}
	\sum_{a,m_a,n_a}
	\Cc^{am_a}_{j_1m_1,\,j_2m_2}\,
	\Cc^{an_a}_{j_1n_1,\,j_2n_2}\,
	\Cc^{jm}_{am_a,\,j_3m_3}\,
	\Cc^{jn}_{an_a,\,j_3n_3} \, .
	\label{eq:app_cg_four_D_integral}
\end{align}
Equations~\eqref{eq:app_cg_three_D_integral} and~\eqref{eq:app_cg_four_D_integral} hold provided $n=n_1+n_2$ and $n=n_1+n_2+n_3$, respectively. These conditions ensure that the coupled product and the conjugated $D$ matrix have the same second index, as required by Eq.~\eqref{eq:app_cg_D_orthogonality}.

The spherical harmonics are realizations of the $\ket{jm}$ basis states (restricted to integer $j$ as we assume in this work) in some direction $\bh{\n}$, $Y_{j m_j}(\bh{\n}) \equiv \braket{\bh{\n}|jm}$. 
The spherical harmonics are related to the Wigner $D$ matrices by
\begin{align}
	Y_{j m_j}(\bh{\n}) 
	& = \sqrt{\frac{2j+1}{4\pi}}\,
	D_{m_j 0}^{(j)\,*}(\bh{\n}) \, .
	\label{eq:app_ylm_to_D}
\end{align}
Thus, the orthogonality relation reads
\begin{align}
	\int\!\diff\bh{\n} \, 
	Y_{jm}^{*}(\bh{\n}) \,
	Y_{j'm'}(\bh{\n})
	= 
	\delta_{j'j} \,
	\delta_{m'm} \, ,
	\label{eq:app_cg_ylm_orthogonality}
\end{align}
and the Clebsch-Gordan series is
\begin{align}
	Y_{j_1m_1}(\bh{\n}) \,
	Y_{j_2m_2}(\bh{\n})
	=
	\sum_{j,m}
	\sqrt{\frac{(2j_1+1)(2j_2+1)}{4\pi(2j+1)}} \, 
	\Cc^{jm}_{j_1m_1,\,j_2m_2}\,
	\Cc^{j0}_{j_10,\,j_20} \,
	Y_{jm}(\bh{\n}) \, .
	\label{eq:app_ylm_cg_series}
\end{align}
Of particular use is the Clebsch-Gordan decomposition,
\begin{align}
	Y_{jm}(\bh{\n}) 
	= 
	\frac{1}{\Cc^{j0}_{j_10,\,j_20}} \,
	\sqrt{\frac{4\pi(2j+1)}{(2j_1+1)(2j_2+1)}}
	\sum_{m_1,m_2}
	\Cc^{jm}_{j_1m_1,\,j_2m_2}\,
	Y_{j_1m_1}(\bh{\n})\,
	Y_{j_2m_2}(\bh{\n})\,.
	\label{eq:app_cg_ylm_decomp}
\end{align}
As with Eq.~\eqref{eq:app_cg_three_D_integral}, the equivalent integral over three spherical harmonics is
\begin{align}
	\int\!\diff\bh{\n}\,
	Y^*_{j m}(\bh{\n})\,
	Y_{j_1m_1}(\bh{\n})\,
	Y_{j_2m_2}(\bh{\n})
	&=
	\sqrt{\frac{(2j_1+1)(2j_2+1)}
	{4\pi(2j+1)}}\,
	\Cc^{j0}_{j_10,\,j_20}
	\Cc^{j m}_{j_1m_1,\,j_2m_2} \, .
	\label{eq:app_cg_gaunt}
\end{align}
This integral is known as the Gaunt coefficient.

Finally, we define the regular solid harmonics by
\begin{align}
	\Yc_{jm}(\n) \equiv n^j\,Y_{jm}(\bh{\n}) \,.
\end{align}
These are homogeneous polynomials of degree $j$ in the Cartesian components of $\n$. 
The main property used in this work is their addition theorem under a finite translation $\n\to\n+\a$,
\begin{align}
	\Yc_{jm}(\n+\a) 
	= 
	\sum_{\ell = 0}^{j} 
	\sqrt{\frac{4\pi}{2\ell+1}\,\binom{2j+1}{2\ell}}\,
	\sum_{\mu = -\ell}^{\ell}
	\Cc^{jm}_{\ell\mu,\, j-\ell,m-\mu}\,
	\Yc_{\ell\mu}(\n) \,
	\Yc_{j-\ell,\,m-\mu}(\a) \, .
	\label{eq:app_cg_solid_harmonic_addition}
\end{align}
%

\section{Structure of the Boost-Coefficient Expansion}
\label{app:sec:boost_coeff}

In Eq.~\eqref{eq:solid_harmonic_expansion}, the boosted solid harmonic contains only spherical harmonics with the same helicity label $\lambda$ and with $\ell\leq S$. 
We prove these two properties here.
We begin with a generic unrestricted orthogonal expansion of the solid harmonic,
\begin{align}
	\Yc_{S\lambda}(\p_k^\star)
	&=
	\sum_{\ell=0}^{\infty}
	\sum_{m=-\ell}^{\ell}
	\Ac^{\ell m}_{S\lambda}(p;k)\,
	Y_{\ell m}(\bh{\p}) \, ,
	\label{eq:app_boost_general_expansion}
\end{align}
where $\p_k^\star=\p_k^\star(\p)$ is obtained by boosting along the quantization axis $-\bh{\k}$, as in Eqs.~\eqref{eq:recap_boost_par} and \eqref{eq:recap_boost_perp}. 
Throughout this appendix, all momenta have their angles defined with respect to this quantization axis; therefore, we do not use the prime ($'$) notation, introduced in Sec.~\ref{sec:derivation}, for convenience.
By the orthogonality of the spherical harmonics, the expansion coefficients are given by
\begin{align}
	\Ac^{\ell m}_{S\lambda}(p;k)
	&=
	\int\!\diff\bh{\p}\,
	Y^*_{\ell m}(\bh{\p})\,
	\Yc_{S\lambda}(\p_k^\star) \, .
	\label{eq:app_boost_general_coeff}
\end{align}
We now show that $\Ac^{\ell m}_{S\lambda}=0$ unless $m=\lambda$ and
$\ell\le S$. 
The $S=0$ case is trivial. 
Considering $S=1$, the Lorentz boost is along
the quantization axis, so the transverse spherical components are unchanged, while the longitudinal component can mix only with the scalar harmonic, \cf~Eqs.~\eqref{eq:boosted_p_spherical_basis}--\eqref{eq:boosted_sph_harm_zero}. 
Thus, the
nonzero terms $\Ac^{\ell m}_{1 \lambda}$ have $m=\lambda$ and $\ell\le1$.

Assume the statement is true through rank $S-1$. 
First, we write the left-hand side of Eq.~\eqref{eq:app_boost_general_expansion} in a Clebsch-Gordan decomposition by applying Eq.~\eqref{eq:app_cg_ylm_decomp} with $j_1=S-1$ and $j_2=1$, and multiplying by
$(\pstark)^S$, which gives
\begin{align}
	\Yc_{S\lambda}(\p_k^\star)
	&=
	\frac{1}{\Cc^{S0}_{S-1,0,\,10}}
	\sqrt{\frac{4\pi(2S+1)}{3(2S-1)}}\,
	\sum_{\Lambda,\mu}
	\Cc^{S\lambda}_{S-1,\Lambda,\,1\mu}\,
	\Yc_{S-1,\Lambda}(\p_k^\star)\,
	\Yc_{1\mu}(\p_k^\star) \, .
	\label{eq:app_boost_induction_start}
\end{align}
Now, applying Eq.~\eqref{eq:app_boost_general_expansion} to the rank-one and rank-($S-1$) solid harmonics, we obtain
\begin{align}
	\Yc_{S\lambda}(\p_k^\star)
	&=
	\frac{1}{\Cc^{S0}_{S-1,0,\,10}}
	\sqrt{\frac{4\pi(2S+1)}{3(2S-1)}}\,
	\sum_{\Lambda,\mu}
	\Cc^{S\lambda}_{S-1,\Lambda,\,1\mu}\nn\\
    &\qquad\qquad \times\sum_{a=0}^\infty 
    \sum_{m_a=-a}^a \Ac^{am_a}_{S-1,\Lambda}(p;k)
	Y_{a,m_a}(\bh{p})\,
    \sum_{b=0}^\infty 
    \sum_{m_b =-b}^b \Ac^{b m_b}_{1\mu}(p;k)
 	Y_{1\mu}(\bh{p}) \, .
	\label{eq:app_boost_induction_pt2}
\end{align}

Using the induction hypothesis and $S=1$ base case result for the $\Ac$ coefficients, this expression may be simplified to
\begin{align}
	\Yc_{S\lambda}(\p_k^\star)
	&=
	\frac{1}{\Cc^{S0}_{S-1,0,\,10}}
	\sqrt{\frac{4\pi(2S+1)}{3(2S-1)}}\,
	\sum_{\Lambda, \mu}
	\Cc^{S\lambda}_{S-1,\Lambda,\,1\mu}
	\nn\\
	&\qquad\qquad\times
	\sum_{a=0}^{S-1}
	\sum_{b=0}^{1}
	\Ac^{a \Lambda}_{S-1,\Lambda}(p;k)
    \Ac^{b \mu}_{1\mu}(p;k)
	Y_{a\Lambda}(\bh{\p})\,
	Y_{b\mu}(\bh{\p}) \, .
	\label{eq:app_boost_induction_insert}
\end{align}
The Clebsch-Gordan coefficient $\Cc^{S\lambda}_{S-1,\Lambda,\,1\mu}$ enforces $\lambda=\Lambda+\mu$. 
The product of spherical harmonics on the right-hand-side of Eq.~\eqref{eq:app_boost_induction_insert} can be combined via the Clebsch-Gordan series, Eq.~\eqref{eq:app_ylm_cg_series}, as
\begin{align}
	Y_{a\Lambda}(\bh{\p}) \,
	Y_{b\mu}(\bh{\p})
	=
	\sum_{\ell,m}
	\sqrt{\frac{(2a+1)(2b+1)}{4\pi(2\ell+1)}} \, 
	\Cc^{\ell m}_{a\Lambda,\,b\mu}\,
	\Cc^{\ell 0}_{a0,\,b0} \,
	Y_{\ell m}(\bh{\p}) \, .
\end{align}
This allows us to write the right-hand side of Eq.~\eqref{eq:app_boost_induction_insert} in the form of Eq.~\eqref{eq:app_boost_general_expansion},
\begin{align}
    \Yc_{S\lambda}(\p_k^\star)
	&= \sum_{\ell, m} \Bigg(\frac{1}{\Cc^{S0}_{S-1,0,\,10}} \sqrt{\frac{ (2S+1)(2a+1)(2b+1)}{3(2S-1)(2\ell+1)}} \sum_{\Lambda, \mu}
	\Cc^{S\lambda}_{S-1,\Lambda,\,1\mu} \nn\\
    &\qquad\qquad\qquad\times 
    \sum_{a=0}^{S-1}
	\sum_{b=0}^{1}
	\Ac^{a \Lambda}_{S-1,\Lambda}(p;k)
    \Ac^{b \mu}_{1\mu}(p;k) \Cc^{\ell m}_{a\Lambda,\,b\mu}\,
	\Cc^{\ell 0}_{a0,\,b0} \,
    \Bigg) Y_{\ell m}(\bh{\p})\,.
    \label{eq:full_exp_unrestricted_coeff}
\end{align}

The Clebsch-Gordan coefficient $\Cc^{\ell m}_{a\Lambda,\,b\mu}$ then enforces
$m=\Lambda+\mu=\lambda$, while the triangle rule gives
$\ell\leq a+b\leq S$. 
Identifying the unrestricted coefficients $\Ac^{\ell m}_{S\lambda}$ of 
Eq.~\eqref{eq:app_boost_general_expansion} with the quantity in the parenthesis of Eq.~\eqref{eq:full_exp_unrestricted_coeff}, we conclude that $\Ac^{\ell m}_{S\lambda}$ vanishes unless $m=\lambda$ and
$\ell\leq S$.
Having concluded our proof, we can compare Eq.~\eqref{eq:app_boost_general_expansion} with Eq.~\eqref{eq:solid_harmonic_expansion} to obtain
\begin{align}
	\Ac^{\ell m}_{S\lambda}(p;k)
	&=
	p^S\,
	\delta_{m\lambda}\,
	\Bc^\ell_{S\lambda}(p;k)
	\, ,
	\label{eq:app_boost_A_to_B}
\end{align}
for $\ell \le S$, and $\Ac^{\ell m}_{S\lambda}=0$ otherwise. 
The factor of $p^S$ is removed by convention so that the boost coefficients are dimensionless. 
Equivalently, the boost coefficients are obtained from Eq.~\eqref{eq:app_boost_general_coeff} as
\begin{align}
	\Bc^\ell_{S\lambda}(p;k)
	&=
	\frac{1}{p^S}
	\int\!\diff\bh{\p}\,
	Y^*_{\ell \lambda}(\bh{\p})\,
	\Yc_{S\lambda}(\p_k^\star) \, ,
	\label{eq:app_boost_B_coeff}
\end{align}
which is the same as Eq.~\eqref{eq:boost_coefficient_def}.

\section{Recoupling of the helicity sum}
\label{app:sec:helicity_recoupling}

In this appendix, we derive the angular-momentum constraints on the helicity sum that appears in the threshold expansion of the vertex function in Sec.~\ref{sec:threshold}. 
The goal is to determine when the low-power terms in Eq.~\eqref{eq:H_threshold} can survive the helicity sum. We begin with
\begin{equation}
    \Sigma^{n,j,J}_{LS}(\ell;\ell')
    \equiv
    \sum_{\lambda}
    \Cc^{S\lambda}_{\ell\lambda,\,S-\ell,0}\,
    \Cc^{J\lambda}_{L0,\,S\lambda}\,
    \Cc^{n\lambda}_{J\lambda,\,j0}\,
    \Cc^{n\lambda}_{\ell\lambda,\,\ell'0} \, .
    \label{app:eq:helicity_sum_definition}
\end{equation}
The range of the helicity sum need not be specified explicitly, since the Clebsch-Gordan coefficients vanish whenever their magnetic quantum numbers lie outside their allowed ranges.
In Sec.~\ref{sec:threshold}, this helicity sum appears in the coefficient of the threshold power $k^{S+\ell'-\ell}$ in the initial-state vertex. 
The analogous final-state sum appears with the power $p^{S'+\ell-\ell'}$, obtained by the interchange $(L,S,\ell,\ell')\to(L',S',\ell',\ell)$. 
The recoupling below shows that any term with too few powers to match the required spin-orbit orbital angular momentum has a vanishing coefficient.

We use the standard Racah recoupling identity
\begin{align}
    &\Cc^{j_{12}m_{12}}_{j_1m_1,\,j_2m_2}\,
    \Cc^{jm_j}_{j_{12}m_{12},\,j_3m_3}
    \nonumber\\
    &\hspace{1cm}
    =
    \sum_{j_{23}}
    (-1)^{j_1+j_2+j_3+j}
    \sqrt{(2j_{12}+1)(2j_{23}+1)}
    \begin{Bmatrix}
        j_1 & j_2 & j_{12} \\
        j_3 & j   & j_{23}
    \end{Bmatrix}
    \Cc^{j_{23}m_{23}}_{j_2m_2,\,j_3m_3}\,
    \Cc^{jm_j}_{j_1m_1,\,j_{23}m_{23}} \, ,
    \label{app:eq:racah_identity}
\end{align}
where $\{\cdots\}$ is the Wigner 6$j$ symbol~\cite{VMK} and $m_{12}$ and $m_{23}$ are fixed by $m_{12} = m_1 + m_2$ and $m_{23} = m_2 + m_3$. 
The explicit expression for the 6$j$ symbol is not needed here, only noting that they are related to recoupling coefficients of three angular momenta.

We start by recoupling the first two Clebsch-Gordan coefficients in Eq.~\eqref{app:eq:helicity_sum_definition}. 
To put them in the order required by Eq.~\eqref{app:eq:racah_identity}, we use the exchange symmetry Eq.~\eqref{eq:app_cg_sym_2} on $\Cc_{L0,\,S\lambda}^{J\lambda}$, giving $\Cc^{J\lambda}_{L0,\,S\lambda} = (-1)^{L+S-J} \,\Cc^{J\lambda}_{S\lambda,\,L0}$, from which it follows that
\begin{align}
    &\Cc^{S\lambda}_{\ell\lambda,\,S-\ell,0}\,
    \Cc^{J\lambda}_{L0,\,S\lambda}
	=
    \sum_x
    \sqrt{(2S+1)(2x+1)}
    \begin{Bmatrix}
        \ell & S-\ell & S \\
        L    & J      & x
    \end{Bmatrix}
    \Cc^{x0}_{S-\ell,0,\,L0}\,
    \Cc^{J\lambda}_{\ell\lambda,\,x0} \,.
    \label{app:eq:first_recoupling}
\end{align}
The phase from Eq.~\eqref{app:eq:racah_identity}, $(-1)^{\ell+(S-\ell)+L+J}$, combines with the exchange phase $(-1)^{L+S-J}$ to give unity, since the exponent is $2(L+S)$.
The remaining pair of Clebsch-Gordans between Eqs.~\eqref{app:eq:helicity_sum_definition} and~\eqref{app:eq:first_recoupling} involving the intermediate angular momentum $J$ can be recoupled in the same way:
\begin{align}
    &\Cc^{J\lambda}_{\ell\lambda,\,x0}\,
    \Cc^{n\lambda}_{J\lambda,\,j0}
	=
    \sum_y
    (-1)^{\ell+x+j+n}
    \sqrt{(2J+1)(2y+1)}
    \begin{Bmatrix}
        \ell & x & J \\
        j    & n & y
    \end{Bmatrix}
    \Cc^{y0}_{x0,\,j0}\,
    \Cc^{n\lambda}_{\ell\lambda,\,y0} \, .
    \label{app:eq:second_recoupling}
\end{align}
Substituting Eqs.~\eqref{app:eq:first_recoupling} and
\eqref{app:eq:second_recoupling} into
Eq.~\eqref{app:eq:helicity_sum_definition} gives
\begin{align}
    \label{app:eq:helicity_sum_before_orthogonality}
    \Sigma^{n,j,J}_{LS}(\ell;\ell')
    ={}&
    \sum_{x,y}
    (-1)^{\ell+x+j+n}
    \sqrt{
        (2S+1)(2x+1)
        (2J+1)(2y+1)
    }
    \nn\\
    &\quad\times
    \begin{Bmatrix}
        \ell & S-\ell & S \\
        L    & J      & x
    \end{Bmatrix}
    \begin{Bmatrix}
        \ell & x & J \\
        j    & n & y
    \end{Bmatrix} \,
    \Cc^{x0}_{S-\ell,0,\,L0}\,
    \Cc^{y0}_{x0,\,j0} \,
    \sum_{\lambda}
    \Cc^{n\lambda}_{\ell\lambda,\,y0}\,
    \Cc^{n\lambda}_{\ell\lambda,\,\ell'0} \,.
\end{align}

The remaining helicity sum is evaluated using the partial orthogonality relation in Eq.~\eqref{eq:partial_orthogonality_1},
\begin{align}
    \sum_{\lambda}
    \Cc^{n\lambda}_{\ell\lambda,\,y0}\,
    \Cc^{n\lambda}_{\ell\lambda,\,\ell'0}
    =
    \frac{2n+1}{2\ell'+1}\,
    \delta_{y,\ell'} \,.
    \label{app:eq:lambda_orthogonality}
\end{align}
Thus, the orthogonality relation fixes the intermediate angular
momentum $y$ to be $\ell'$. 
Performing the $y$ sum in
Eq.~\eqref{app:eq:helicity_sum_before_orthogonality}, we obtain
\begin{align}
    \label{app:eq:fully_recoupled_helicity_sum}
    \Sigma^{n,j,J}_{LS}(\ell;\ell')
    ={}&
    (2n+1)
    \sqrt{
        \frac{(2S+1)(2J+1)}
             {2\ell'+1}
    }
    \sum_x
    (-1)^{\ell+x+j+n}
    \sqrt{2x+1}
    \nn\\
    &\quad \times
    \begin{Bmatrix}
        \ell & S-\ell & S \\
        L    & J      & x
    \end{Bmatrix}
    \begin{Bmatrix}
        \ell & x & J \\
        j    & n & \ell'
    \end{Bmatrix} \,
    \Cc^{x0}_{S-\ell,0,\,L0}\,
    \Cc^{\ell'0}_{x0,\,j0} \, .
\end{align}

Equation~\eqref{app:eq:fully_recoupled_helicity_sum} exposes the relevant selection rule. 
A nonzero term in the $x$ sum requires both
\begin{align}
    \left\lvert L-(S-\ell)\right\rvert
    \le x \le
    L+S-\ell,
    \,\,\, \textrm{and }
    \left\lvert j-\ell'\right\rvert
    \le x \le
    j+\ell' \, .
    \label{app:eq:x_triangle_conditions}
\end{align}
In particular, the first triangle relation implies
$L\le (S-\ell)+x$. 
Combining this with the other triangle relation gives
\begin{align}
    L
    \le
    (S-\ell)+x
    \le
    (S-\ell)+j+\ell' \, ,
\end{align}
and hence
\begin{align}
    \ell'+S-\ell
    \ge
    L-j \, .
    \label{app:eq:initial_weaker_bound}
\end{align}

The same triangle conditions also imply the complementary inequality. 
Indeed, $j\le x+\ell'$ and $x\le L+S-\ell$, so that
\begin{align}
    j
    \le
    x+\ell'
    \le
    L+S-\ell+\ell' \, .
\end{align}
Combining this relation with Eq.~\eqref{app:eq:initial_weaker_bound} gives the stronger condition
\begin{align}
    \ell'+S-\ell
    \ge
    \left\lvert L-j\right\rvert \,.
    \label{app:eq:initial_stronger_bound}
\end{align}
If Eq.~\eqref{app:eq:initial_stronger_bound} fails, there is no value of $x$ for which both Clebsch-Gordan coefficients in Eq.~\eqref{app:eq:fully_recoupled_helicity_sum} are nonzero, and the coefficient of the corresponding low-power term therefore vanishes identically. 
The final-state constraint follows by the replacements
$(L,S,\ell,\ell')\to(L',S',\ell',\ell)$. 
Thus the corresponding low-power term can be nonzero only if
\begin{align}
    \ell+S'-\ell'
    \ge
    \left\lvert L'-j\right\rvert \, .
    \label{app:eq:final_stronger_bound}
\end{align}
%

\section{Partial-wave OPE amplitudes as a function of \texorpdfstring{$s$}{s}}
\label{app:sec:plots_in_s}
\begin{figure}[t]
    \centering
    \includegraphics[width=\linewidth]{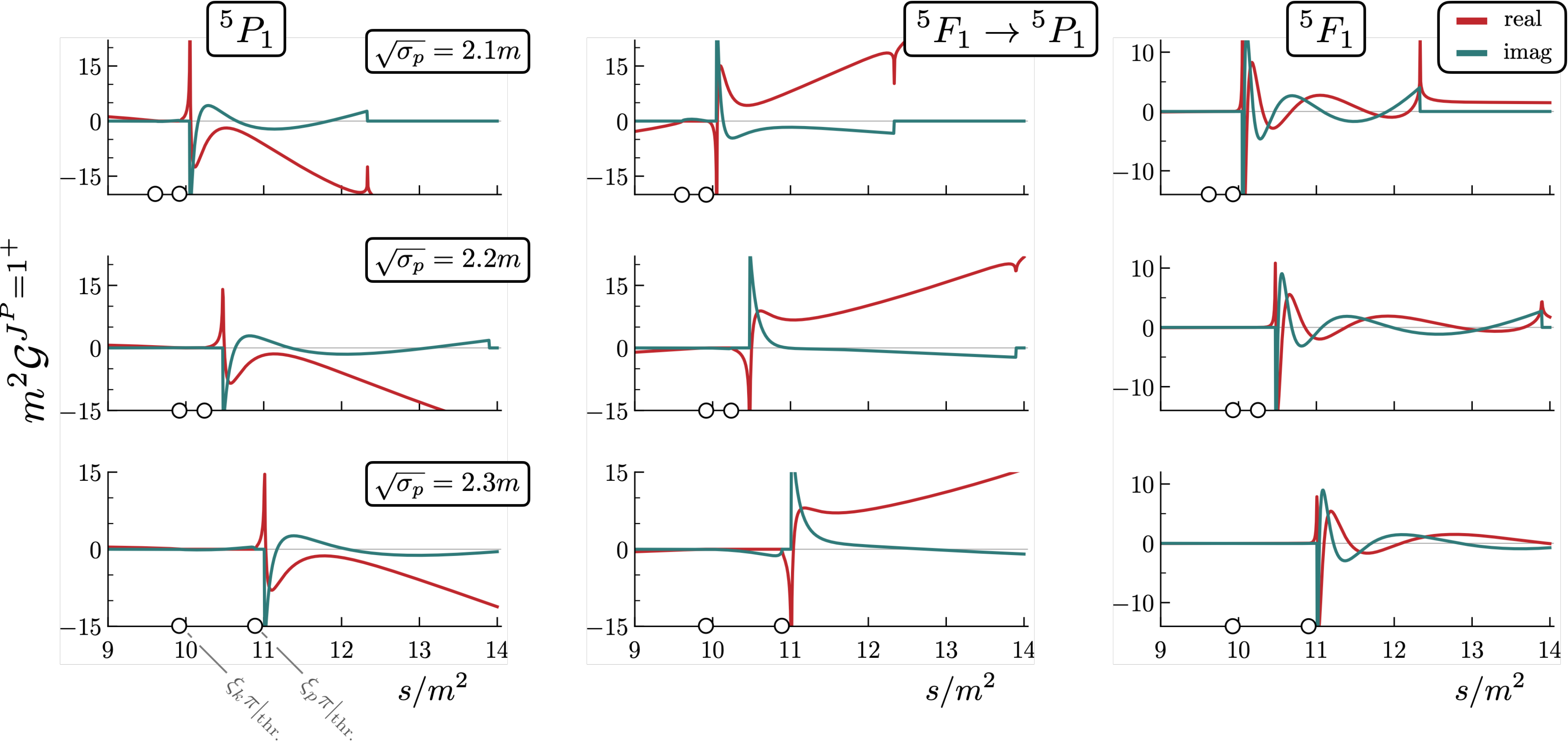}
    \caption{Real and imaginary parts of the OPE amplitude for $J^P = 1^+$, plotted as a function of $s/m^2$ in the region $9 \leq s/m^2 \leq 14$. 
    As in Figs.~\ref{fig:gjp_1p_sigma} through \ref{fig:gjp_3m_sigma}, we fix $\sqrt{\sigma_k}/m = 2.15$ and vary the remaining variable $\sqrt{\sigma_p}/m = \{2.1, 2.2, 2.3\}$ across the rows of the figure. 
    The same spin-orbit quantum number configurations are given in each column as in Fig.~\ref{fig:gjp_1p_sigma}. 
    Marked along the horizontal axis of each plot are the outgoing and incoming pair-spectator production thresholds, $\xi_p \pi\vert_{\textrm{thr.}} = (\sqrt \sigma_p + m)^2$ and similarly for $\xi_k \pi\vert_{\textrm{thr.}}$.}
    \label{fig:gjp_1p_s}
\end{figure}
\begin{figure}[t]
    \centering
    \includegraphics[width=\linewidth]{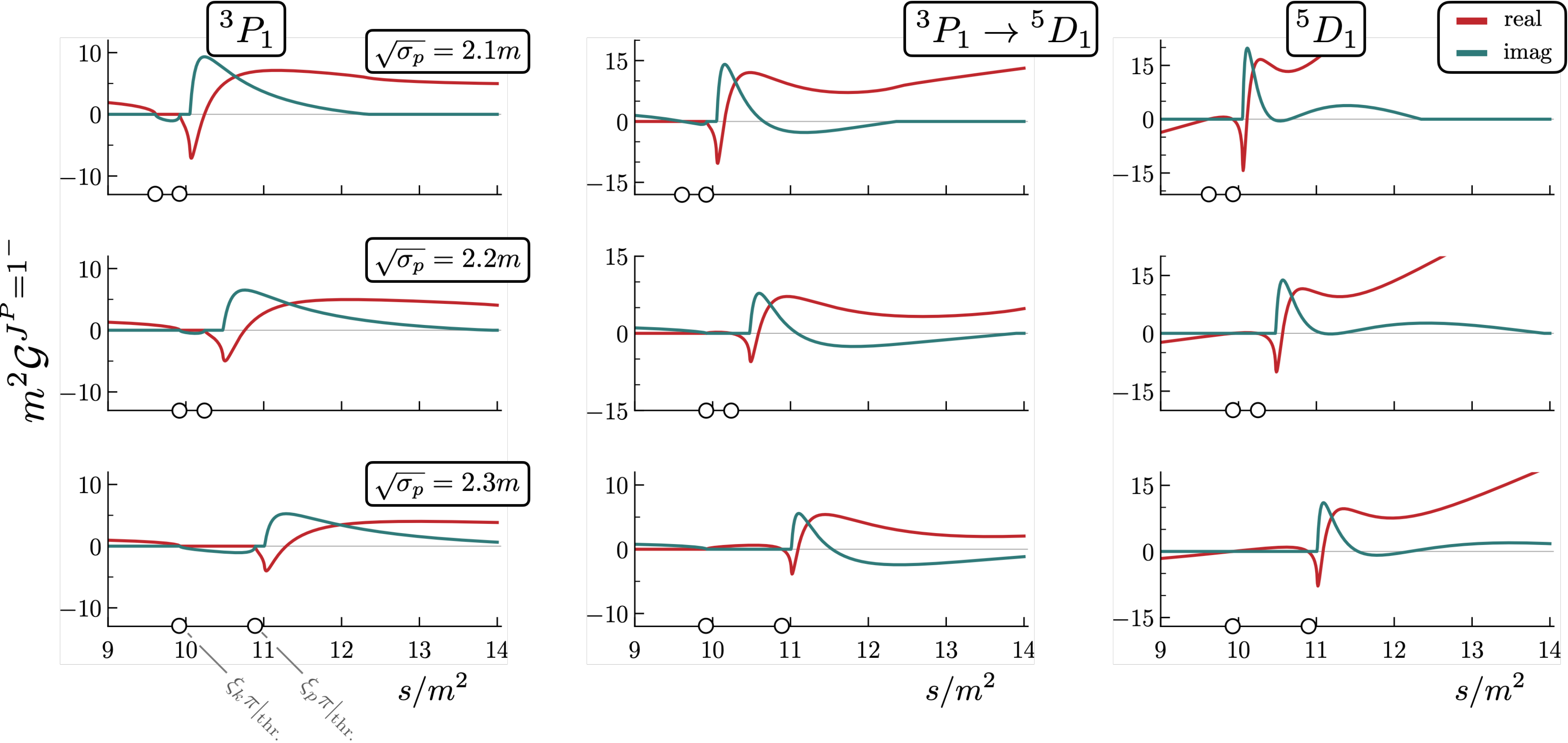}
    \caption{Same as Fig.~\ref{fig:gjp_1p_s}, but for $J^P = 1^-$.}
    \label{fig:gjp_1m_s}
\end{figure}

\begin{figure}
    \centering
    \includegraphics[width=\linewidth]{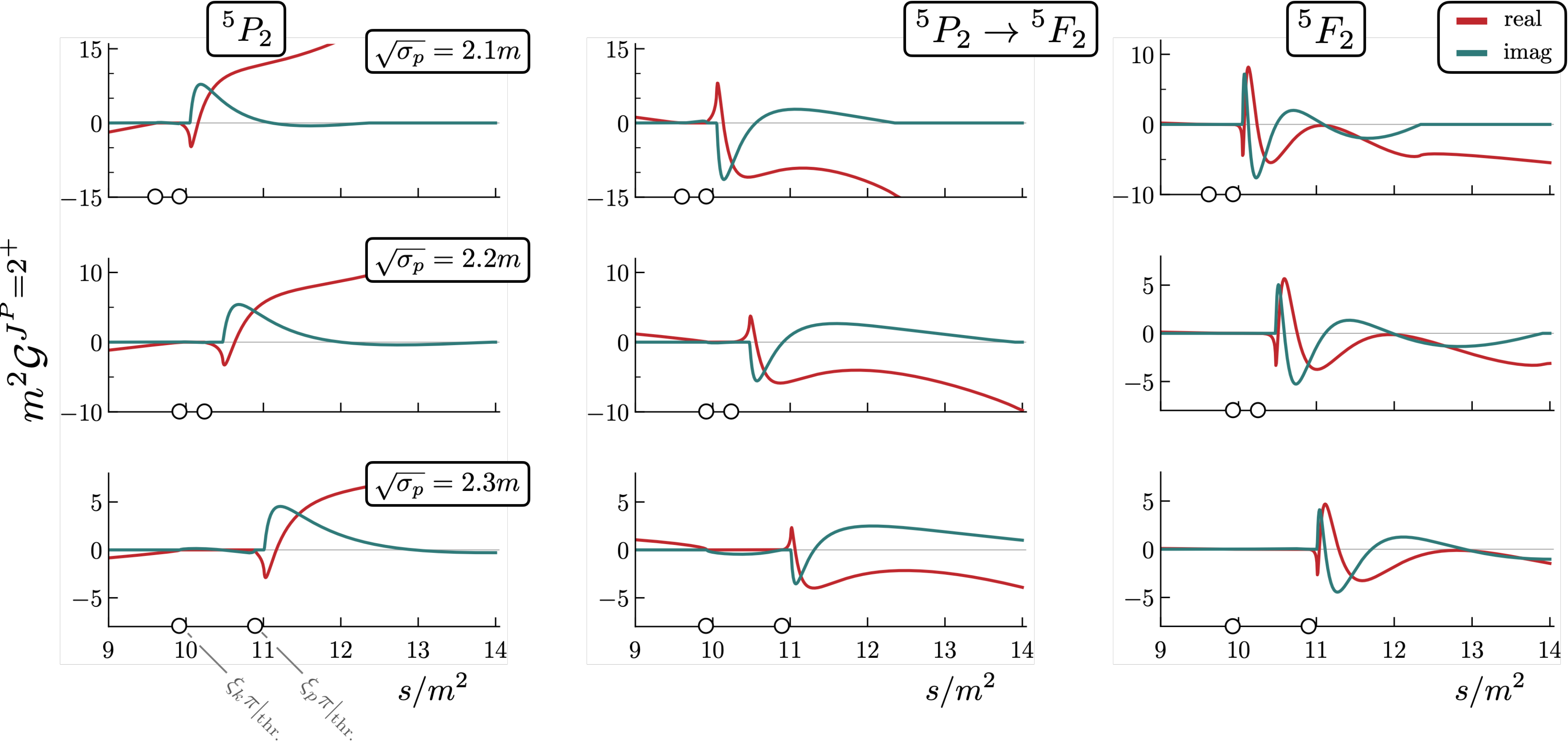}
    \caption{Same as Fig.~\ref{fig:gjp_1p_s}, but for $J^P = 2^+$.}
    \label{fig:gjp_2p_s}
\end{figure}

\begin{figure}
    \includegraphics[width=0.45\textwidth]{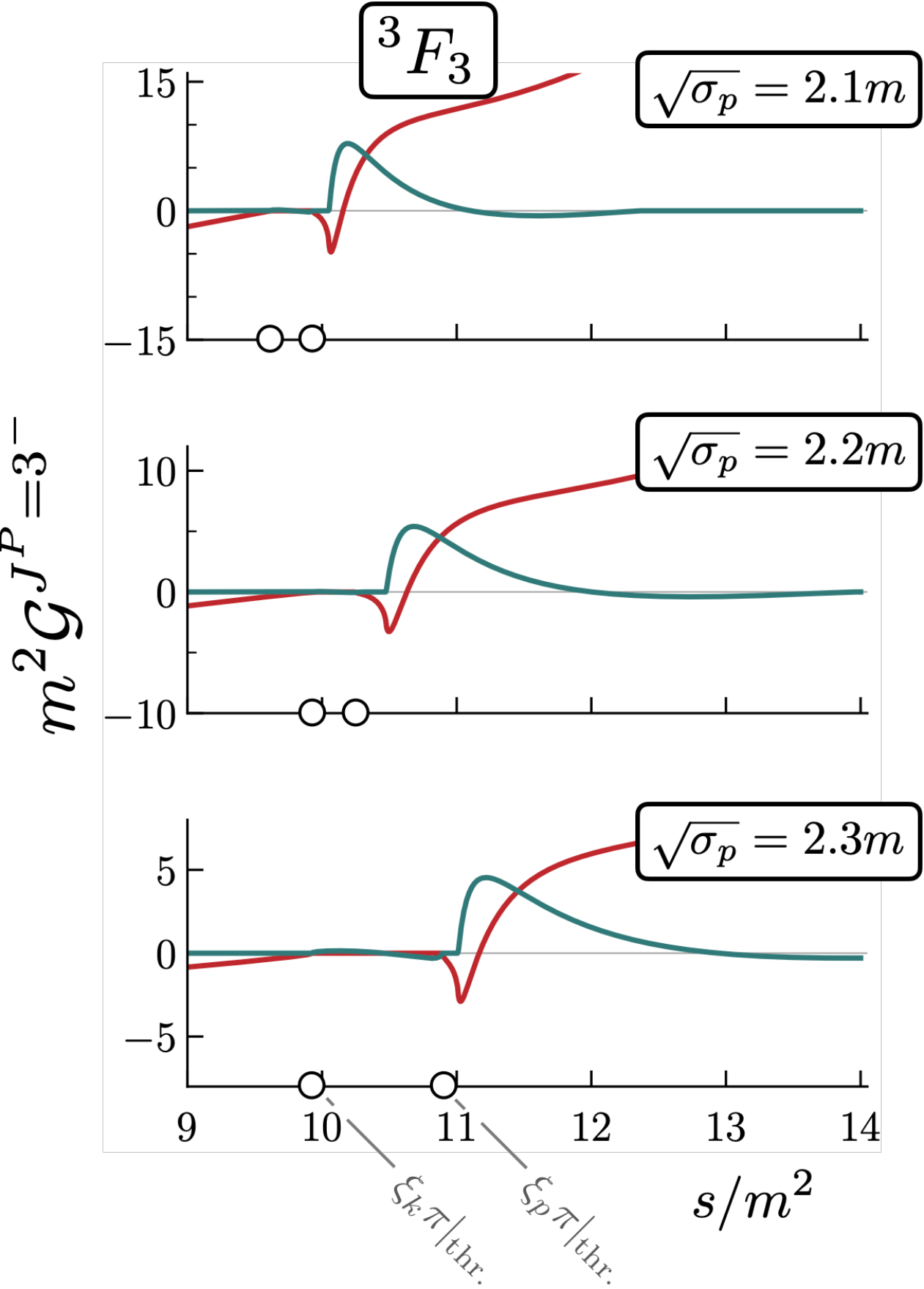}
    \caption{Same as Fig.~\ref{fig:gjp_1p_s}, but for $J^P = 3^-$.}
    \label{fig:gjp_3m_s}
\end{figure}

We further visualize the same OPE functions plotted in Figs.~\ref{fig:gjp_1p_sigma}--\ref{fig:gjp_3m_sigma}, now plotted as functions of $s/m^2$ in the region $9 \leq s/m^2 \leq 14$. 
In Figs.~\ref{fig:gjp_1p_s} through \ref{fig:gjp_3m_s}, we fix $\sqrt{\sigma_k}/m = 2.15$ and vary $\sqrt{\sigma_p}/m = \{2.1, 2.2, 2.3\}$ by row. 
Each column contains the same spin-orbit quantum number configurations contributing to the target $J^P$ in an identical manner to Figs.~\ref{fig:gjp_1p_sigma}--\ref{fig:gjp_3m_sigma}. 
As in Ref.~\cite{Jackura:2023qtp}, we denote the quasiparticle pair as $\xi$, with a $p$ or $k$ subscript to indicate which spectator it scatters against.
Marked on each plot are the threshold values of the effective quasiparticle-spectator scattering systems, $\xi_k \pi$ and $\xi_p \pi$. 
At fixed values of $\sigma_p$ and $\sigma_k$, these threshold values in $s$ are given by $\xi_k \pi\vert_{\textrm{thr.}} = (\sqrt{\sigma_k} + m)^2$ and $\xi_p \pi\vert_{\textrm{thr.}} = (\sqrt{\sigma_p} + m)^2$. 
The larger of these two thresholds defines the value in $s$ at which we have the expected momentum scaling $k^L$ or $p^{L'}$ that was illustrated in Fig.~\ref{fig:ope_threshold_dep}.
For $J^P = 1^+$ shown in Fig.~\ref{fig:gjp_1p_s}, we observe some features of the OPE function that are consistent across the remaining cases. 
First, we have branch points in $s$, just as in $\sigma_p$ in Figs.~\ref{fig:gjp_1p_sigma}--\ref{fig:gjp_3m_sigma}, that are associated with the on-shell points of physical OPE. 
The branch points locations, $s^{-}$ and $s^{+}$ are given by
\begin{align}
    s^{\pm} &= \frac{1}{2m^2}\big[ (\sigma_k - m^2)(\sigma_p - m^2) + m^2 (\sigma_p + \sigma_k + m^2) \nonumber\\
    &\qquad\qquad\pm \lambda^{1/2}(m^2, m^2, \sigma_k) \lambda^{1/2}(m^2, m^2, \sigma_p)\big] \,.
\end{align}
As $\sigma_p$ increases, the region of physical OPE increases, thereby widening the distance in $s$ between the branch points. 
Above the highest threshold, the imaginary part of the OPE takes a nonzero value only in the region where OPE can occur on-shell, consistent with unitarity.

In Fig.~\ref{fig:gjp_1m_s}, we have some examples coinciding with $J^P = 1^-$. 
As in the corresponding plots in $\sigma_p$, we see a general decrease in the number of critical points in the physical OPE region, which is further discussed in App.~\ref{app:sec:parity_suppression}. 
We also see a suppression of the OPE singularities for the same reasons as in the plots in $\sigma_p$. 
In Figs.~\ref{fig:gjp_2p_s} and \ref{fig:gjp_3m_s}, we again have plots of the OPE function for higher spin $J^P = 2^+$ and $3^-$.

\section{Parity suppression of the highest Legendre degree}
\label{app:sec:parity_suppression}

In the numerical examples of Sec.~\ref{sec:numerical}, certain parity sectors exhibit fewer critical points in the physical OPE region than their opposite-parity counterparts. 
This behavior can be traced to a simple helicity symmetry. For some $J^P$ sectors, the highest allowed Legendre degree, $j=\jmaxcg=J+S+S'$, vanishes after transforming from the helicity basis to the spin-orbit basis.

We show this using the spin-orbit recoupling of the angular coefficient,
\begin{align}
	\Hc^{j,J^P}_{L'S',\,LS}(p,k)
	&=
	\sum_{\lambda'=-\lambda'_{\max}}^{\lambda'_{\max}}
	\sum_{\lambda=-\lambda_{\max}}^{\lambda_{\max}}
	\Pc_{\lambda'}({}^{2S'+1}L'_J)\,
	\Hc^{j,J}_{S'\lambda',\,S\lambda}(p,k)\,
	\Pc_{\lambda}({}^{2S+1}L_J) \, .
	\label{eq:app_H_coeff_so}
\end{align}
The spin-orbit coupling coefficient satisfies
\begin{align}
    \Pc_{-\lambda}({}^{2S+1}L_J)
    =
    (-1)^{L+S-J}\Pc_{\lambda}({}^{2S+1}L_J) \, ,
\end{align}
which follows from its definition in terms of Clebsch-Gordan coefficients, Eq.~\eqref{eq:spin_orbit_coupling_def}, and their subsequent symmetry relation Eq.~\eqref{eq:app_cg_sym_1}.

We next examine the helicity coefficient $\Hc_{S'\lambda',\,S\lambda}^{j,J}$ under the helicity flip $\lambda\to-\lambda$. 
For this purpose it is useful to repeat Eq.~\eqref{eq:H_helicity_final},
\begin{align}
	\Hc^{j,J}_{S'\lambda',\,S\lambda}(p,k) 
	& = 
	\sum_{\ell'=0}^{S'} 
	\sum_{\ell=0}^{S} 
	\sum_{n = \lvert\ell'-\ell\rvert}^{\ell'+\ell}
	\frac{(-1)^{n}}{2n+1}
	\left(\frac{k}{q_p^\star}\right)^{S'}
	\sqrt{2\ell'+1}\,
	\Bc_{S'\lambda'}^{\ell'}(k;p) \,
	\Cc_{J\lambda',\,j0}^{n\lambda'} \,
	\Cc_{\ell'\lambda',\,\ell0}^{n\lambda'}
	\nn \\[5pt]
	& \qquad \times
	\left(\frac{p}{q_k^\star}\right)^{S}
	\sqrt{2\ell+1}\,
	\Bc_{S\lambda}^{\ell}(p;k) \,
	\Cc_{J\lambda,\,j0}^{n\lambda} \,
	\Cc_{\ell\lambda,\,\ell'0}^{n\lambda} \, .
	\label{eq:app_H_coeff_hel}
\end{align}
By inspection through the recursive formula Eq.~\eqref{eq:boost_coeff_recursive}, the boost coefficients are invariant under this sign flip. 
Thus the $\lambda\to -\lambda$ behavior is controlled by the two Clebsch-Gordan coefficients in Eq.~\eqref{eq:app_H_coeff_hel}. 
Using the symmetry relations of App.~\ref{app:sec:ang_mom}, one finds 
\begin{align}
    \label{eq:app_cg_symm}
    \Cc^{n,-\lambda}_{J,-\lambda,\,j0}
    \Cc^{n,-\lambda}_{\ell,-\lambda,\,\ell'0}
    =
    (-1)^{J+j+\ell+\ell'}
    \Cc^{n\lambda}_{J\lambda,\,j0}
    \Cc^{n\lambda}_{\ell\lambda,\,\ell'0} \,.
\end{align}
For the highest allowed Legendre degree, $j=\jmaxcg=J+S+S'$, the only nonvanishing terms have $\ell+\ell'=S+S'$. 
The phase in Eq.~\eqref{eq:app_cg_symm} is then
$(-1)^{2(J+S+S')}=1$, and therefore we find
\begin{align}
	\Hc^{j,J}_{S'\lambda',\,S,{-\lambda}}(p,k)
	=
	\Hc^{j,J}_{S'\lambda',\,S\lambda}(p,k) \, , 
\end{align}
for $j=\jmaxcg$. 
An analogous relation holds under $\lambda'\to-\lambda'$.

Thus the highest-$j$ helicity coefficient is even under helicity sign flips. In the spin-orbit sum, Eq.~\eqref{eq:app_H_coeff_so}, it is multiplied by $\Pc_\lambda$, which is odd under $\lambda\to-\lambda$ whenever $L+S-J$ is odd. 
In that case the helicity sum cancels pairwise, and
$\Hc^{j,J^P}_{L'S',\,LS}$ vanishes at $j=\jmaxcg$. 
For three pseudoscalars, this condition corresponds to $P=(-1)^J$. 
The leading Legendre degree is therefore suppressed in these parity sectors, explaining the reduced critical-point structure observed in Fig.~\ref{fig:gjp_1m_sigma} relative to Fig.~\ref{fig:gjp_1p_sigma}.

\bibliography{master.bib}

\end{document}